\documentclass[preprint,pre]{revtex4-1}
\usepackage{amssymb,amsfonts,amsmath,bm,mathrsfs,graphicx,color,subfigure}

\begin{document}
\title{Optical Theorem in Nonlinear Media}
\author{Wei Li}
\affiliation{Department of Mathematics, University of Michigan, Ann Arbor, MI 48109}
\email{leewei@umich.edu}

\author{John C. Schotland}
\affiliation{Department of Mathematics and Department of Physics, University of Michigan, Ann Arbor, MI 48109}
\email{schotland@umich.edu}

\date{\today}

\begin{abstract}
We derive the optical theorem for scattering of electromagnetic waves in nonlinear media. This result is used to obtain the power extinguished from a field by a nonlinear scatterer. The cases of second harmonic generation and the Kerr effect are studied in some detail. Applications to nonlinear apertureless scanning near-field optical microscopy are considered.
\end{abstract}

\maketitle

\section{Introduction}
{
The optical theorem is a basic result in scattering theory that is of both fundamental interest and considerable applied importance~\cite{Newton}. 
It can be formulated in a variety of settings, including quantum mechanics, acoustics and electromagnetic theory. In its simplest form, the optical theorem relates the power extinguished from a plane wave incident on a scattering medium to the scattering amplitude in the forward direction of scattering. For the case of electromagnetic scattering, the extinguished power $P_e$ is given by 
\begin{equation}
\label{opt thm}
P_e=\frac{c}{2k_0}\text{Im}\textbf{A}\cdot\textbf{E}^*_{0}  \ ,
\end{equation}
where $k_0$ is the wavenumber, $\textbf{E}_{0}$ is the incident field and $\textbf A$ is the scattering amplitude in the forward direction~\cite{van_de_Hulst,Jones,Born-Wolf}. In physical terms, the loss of power from the incident field is due to interference between the incident field and the scattered field within the volume of the scatterer.}

{
The standard derivation of the optical theorem makes use of the ansatz that the scattered field $\textbf{E}_s$ 
behaves asymptotically as an outgoing spherical wave of the form
\begin{equation}
\label{ansatz}
{\bf{E}}_s \sim {\bf{A}} \frac{e^{ik_0r}}{r} \ , \quad k_0r \to\infty \ .
\end{equation}
The above ansatz may be justified for the case of material media
in which the polarization  is related to the electric field by a \emph{linear} constitutive relation. Moreover, in this situation, the optical theorem may be derived without invoking the asymptotic behavior of the scattered field~\cite{Carney,Lytle}.} 

{
The optical theorem is normally considered within the framework of linear optics~\cite{Born-Wolf}. However, the ansatz (\ref{ansatz}) is very general.
That is, all properties of the scatterer are encoded in its scattering amplitude, which can, in principle, be arbitrarily prescribed. Thus, there is no reason to restrict the optical theorem to the linear response regime. Making use of this observation, in this paper we consider the optical theorem in the 
context of nonlinear media. We derive an expression for the extinguished power that holds when the polarization is an arbitrary function of the electric field. To some extent, this result may be expected on physical grounds. However, we provide a proper mathematical justification following the approach of~\cite{Carney,Lytle}. We specialize our result to the cases of quadratic and cubic nonlinearities. We also study in detail the processes of second harmonic generation and the Kerr effect for small scatterers. Our results on scattering from small nonlinear particles are of independent interest, since exact solutions to nonlinear scattering problems are, to the best of our knowledge, known only in one dimension~\cite{Shen,Boyd}.} 

The remainder of this paper is organized as follows. In section II, we present some basic results in nonlinear optics and scattering theory that will be used later in the paper. Section III presents the derivation of the optical theorem in the form we require, without the use of asymptotics. In section IV, we consider separately the cases of second- and third-order nonlinearities. Numerical results for small scatterers are presented in section V. Finally, in section VI we consider applications to apertureless scanning near-field optical microscopy in which the tip exhibits a nonlinear optical response. We investigate the cases of second- and third-harmonic generation and characterize the achievable resolution for a model system consisting of two scatterers. A discussion of our results is presented in Section VII. 

\section{Preliminaries}
In this section we collect several results in nonlinear optics and scattering theory that
will be useful in the derivation of the generalized optical theorem. We begin by recalling that
the Maxwell equations in a source-free nonmagnetic medium are of the form
\begin{eqnarray}
\label{Maxwell_1}
\nabla\cdot\textbf{D}&=&0 \ , \\
\label{Maxwell_2}
\nabla\times\textbf{E}+\frac{1}{c}\frac{\partial\textbf{B}}{\partial t}&=&0 \  ,  \\
\label{Maxwell_3}
\nabla\cdot\textbf{B}&=&0 \ ,  \\
\label{Maxwell_4}
\nabla\times\textbf{B}-\frac{1}{c}\frac{\partial\textbf{D}}{\partial t}&=&0 \ .
\end{eqnarray}
Here $\textbf{E}$ is the electric field, $\textbf{B}$ is the magnetic field, $\textbf{D}$ is the electric displacement field and $\textbf{P}$ is the polarization. In addition, $\textbf{D}$ and $\textbf{P}$ satisfy the relation
\begin{eqnarray}
\textbf{D}&=&\textbf{E}+4\pi\textbf{P}\ .
\end{eqnarray}

Throughout this paper, we will  use the following  Fourier transformation convention: 
\begin{eqnarray}
f(\textbf{r},\omega) &=&\int f(\textbf{r},t)e^{i\omega t}dt \ , \\
f(\textbf{r},t) &=&\frac{1}{2\pi}\int f(\textbf{r},\omega)e^{-i\omega t}d\omega \  ,
\end{eqnarray}
where the time and frequency dependence of all quantities are displayed explicitly.
We note that if $f(\textbf{r},t)$ is real-valued, then $f(\textbf{r},-\omega)=f^*(\textbf{r},\omega)$.
Upon Fourier transforming (\ref{Maxwell_1})--(\ref{Maxwell_4}) we obtain
\begin{eqnarray}
\label{Maxwell f}
\nabla\cdot\textbf{D}(\textbf{r},\omega)&=&0\ , \\
\nabla\times\textbf{E}(\textbf{r},\omega)-ik(\omega)\textbf{B}(\textbf{r},\omega)&=&0\ , \\
\nabla\cdot\textbf{B}(\textbf{r},\omega)&=&0\ , \\
\nabla\times\textbf{B}(\textbf{r},\omega)+ik(\omega)\textbf{D} (\textbf{r},\omega) &=&0\ ,
\end{eqnarray}
where $k(\omega)=\omega/c$. 
%

If the medium is linear, the polarization is given by
\begin{eqnarray}
\label{linear polarization}
P_i(\textbf{r},\omega)=\chi^{(1)}_{ij}(\textbf{r};\omega)E_j(\textbf{r},\omega)\ ,
\end{eqnarray}
where $\chi^{(1)}_{ij}$ is the first-order susceptibility. Here we have adopted the summation convention, whereby repeated indices are summed. For the case of quadratic nonlinear media,
\begin{eqnarray}
\label{quadratic polarization}
P_i(\textbf{r},\omega)=\chi^{(1)}_{ij}(\textbf{r};\omega)E_j(\omega)+\sum_{\omega_1+\omega_2=\omega}\chi^{(2)}_{ijk}(\textbf{r};\omega_1,\omega_2)E_j(\textbf{r},\omega_1)E_k(\textbf{r},\omega_2)\ ,
\end{eqnarray}
where $\chi^{(2)}_{ijk}$ is the second-order susceptibility. The sum implies that  the electric field at the frequencies $\omega_1$ and $\omega_2$ contributes to the polarization at the frequency $\omega$ if $\omega_1+\omega_2=\omega$. For cubic nonlinear media,
\begin{eqnarray}
\label{cubic polarization}
P_i(\textbf{r},\omega)=\chi^{(1)}_{ij}(\textbf{r};\omega)E_j(\omega)+\sum_{\omega_1+\omega_2+\omega_3=\omega}\chi^{(3)}_{ijkl}(\textbf{r};\omega_1,\omega_2,\omega_3)E_j(\textbf{r},\omega_1)E_k(\textbf{r},\omega_2)E_l(\textbf{r},\omega_3)\ ,
\end{eqnarray}
where $\chi^{(3)}_{ijkl}(\textbf{r};\omega_1,\omega_2,\omega_3)$ are the third order susceptibilities. 

We will assume that the susceptibilities have over all permutation symmetry.
This assumption is quite standard and holds for nonresonance frequencies in the classical anharmonic oscillator model and in quantum optics~\cite{Shen,Boyd}. However, we note that Kleinman symmetry can be broken in a variety of systems~\cite{Dailey,Xu,Tsutsumi,Shelton,Singh,Chemla}.

The wave equation for the electric field $\textbf{E}(\textbf{r},\omega)$ is obtained by
taking the curl of (\ref{Maxwell_2}) and eliminating the magnetic field $ \textbf{B}(\textbf{r},\omega)$
using (\ref{Maxwell_4}).
We then have
\begin{eqnarray}
\label{wave eqn}
\nabla\times\nabla\times\textbf{E}(\textbf{r},\omega)-k^2(\omega)\textbf{E}(\textbf{r},\omega)=4\pi k^2(\omega)\textbf{P}(\textbf{r},\omega)\ .
\end{eqnarray}
The solution of the wave equation (\ref{wave eqn}) is given by
\begin{eqnarray}
\label{soln}
E_{i}(\textbf{r},\omega)=E_{inc,i}(\textbf{r},\omega) + k^2(\omega)\int d^3 r' G_{ij}(\textbf{r},\textbf{r}';\omega)P_j(\textbf{r}',\omega)\ ,
\end{eqnarray}
where $\textbf{E}_{inc}(\textbf{r},\omega)$ obeys (\ref{wave eqn}) with $P=0$.
The Green's function $G_{ij}$ is of the form~\cite{Tai}
\begin{eqnarray}
G_{ij}(\textbf{r},\textbf{r}';\omega)=\left(\delta_{ij}+\frac{1}{k^2(\omega)}\partial_i\partial_j\right)G(\textbf{r},\textbf{r}';\omega)\  ,
\end{eqnarray}
where
\begin{eqnarray}
G(\textbf{r},\textbf{r}';\omega)=\frac{e^{ik(\omega)|\textbf{r}-\textbf{r}'|}}{|\textbf{r}-\textbf{r}'|} 
\ .
\end{eqnarray}
Straightforward calculation shows that the Green's function has the following asymptotic form 
\begin{eqnarray}
\label{G asymp}
G_{ij}(\textbf{r},\textbf{r}';\omega)\sim\frac{e^{ik(\omega)r}}{r}(\delta_{ij}-\hat{r}_i\hat{r}_j)e^{-ik(\omega)\hat{\textbf{r}}\cdot\textbf{r}'}\ ,
\end{eqnarray}
when $r \gg r'$. Using this result, we find that the scattered field behaves as an outgoing spherical wave of the form
\begin{eqnarray}
E_{s,i}(\textbf{r},\omega)\sim\frac{e^{ik(\omega)r}}{r}A_i(\hat{\textbf{r}},\omega) \  ,
\end{eqnarray}
where the scattering amplitude is defined by
\begin{eqnarray}
\label{scattering amp}
A_i(\hat{\textbf{r}},\omega)=k^2(\omega)(\delta_{ij}-\hat{r}_i\hat{r}_j)\int d^3 r' P_j(\textbf{r}',\omega)e^{ik(\omega)\hat{\textbf{r}}\cdot\textbf{r}'}\ .
\end{eqnarray}

Following standard procedures, the conservation of energy follows immediately from (\ref{wave eqn}) and takes the form
\begin{eqnarray}
\label{div current}
\nabla\cdot\textbf{S}(\textbf{r},\omega)=\frac{ck(\omega)}{2\pi^2}\text{Im}(\textbf{E}^*(\textbf{r},\omega)\cdot\textbf{P}(\textbf{r},\omega))\  ,
\end{eqnarray}
where the Poynting vector $\textbf{S}$ is defined by
\begin{eqnarray}
\label{Poynting f}
\textbf{S}(\textbf{r},\omega)=\frac{c}{8\pi^3}\text{Re}(\textbf{E}(\textbf{r},\omega)\times\textbf{B}^*(\textbf{r},\omega)) \ .
\end{eqnarray}

We recall that the time-dependent Poynting vector is defined as
\begin{eqnarray}
\textbf{S}(\textbf{r},t)=\frac{c}{4\pi}\textbf{E}(\textbf{r},t)\times\textbf{B}(\textbf{r},t) \ .
\end{eqnarray}
For time-harmonic fields, the time average of the Poynting vector
\begin{eqnarray}
\label{time averaged Poynting vector}
\bar{\textbf{S}} = \lim_{T\rightarrow \infty}\frac{1}{T}\int_0^T\textbf{S}(\textbf{r},t)dt 
\end{eqnarray} 
is well defined. It follows that
\begin{eqnarray}
\label{Poynting t}
\bar{\textbf{S}}=\int_{0}^{\infty}\textbf{S}(\textbf{r},\omega) d\omega \ .
\end{eqnarray} 
%

We note that for nondispersive media,  the analog of the Manley-Rowe relations can  be shown to hold. That is, if the susceptibilities $\chi_{ij}^{(1)}$, $\chi_{ijk}^{(2)}$ and $\chi_{ijkl}^{(3)}$ are purely real, then
$\nabla \cdot \bar{\textbf{S}} =0$. The proof is given in Appendix A.

\section{Optical Theorem}
In this section we derive the optical theorem for nonlinear media following the approach of~\cite{Carney,Lytle}. We begin by recalling some basic facts from scattering theory~\cite{Born-Wolf}. We consider a general nonlinear medium, whose
polarization is defined by either a quadratic or cubic nonlinearity. We suppose that a field $\textbf{E}_{inc}$ is incident upon the medium and write the total electric field as the sum 
\begin{equation}
\textbf{E} = \textbf{E}_{inc}  + \textbf{E}_s  \ ,
\end{equation}
where $\textbf{E}_{s}$ is the scattered field. 
It follows from (\ref{wave eqn}) that the scattered field obeys
\begin{eqnarray}
\label{wave eqn s}
&&\nabla\times\nabla\times\textbf{E}_s(\textbf{r},\omega)-k^2(\omega)\textbf{E}_s(\textbf{r},\omega)=4\pi k^2(\omega)\textbf{P}(\textbf{r},\omega) \ .
\end{eqnarray}
The energy carried by the scattered field is associated with the Poynting vector 
$\textbf{S}_s$, which is defined by
\begin{eqnarray}
\textbf{S}_s=\frac{c}{8\pi^3}\text{Re}(\textbf{E}_s(\omega)\times\textbf{B}_s^*(\omega)) \ .
\end{eqnarray}
Evidently, $\textbf{S}_s$ obeys the conservation law
\begin{eqnarray}
\label{div current s}
\nabla\cdot\textbf{S}_s(\textbf{r},\omega)=\frac{ck(\omega)}{2\pi^2}\text{Im}(\textbf{E}_s^*(\textbf{r},\omega)\cdot\textbf{P}(\textbf{r},\omega))\  ,
\end{eqnarray}
which is a consequence of (\ref{wave eqn s}). 

Suppose that the scattering medium is contained in a volume $V$. Then the power absorbed by the medium is given by
\begin{eqnarray}
P_a(\omega)=-\int_{\partial V}\textbf{S}(\textbf{r},\omega)\cdot\hat{\textbf{n}}d^2r\ ,
\end{eqnarray}
where $\hat{\textbf{n}}$ is the outward unit normal to $\partial V$ and the presence of the overall minus sign signifies that this is the flux of the Poynting vector of the wave traveling into the medium.
Converting the above surface integral to a volume integral by means of the divergence theorem and making use of (\ref{div current}), we have
\begin{eqnarray}
\label{Pa}
P_a(\omega)=\frac{ck(\omega)}{2\pi^2}\int_V \text{Im}(\textbf{E}^*(\textbf{r}',\omega)\cdot\textbf{P}(\textbf{r}',\omega))d^3r' \ .
\end{eqnarray}
In a strictly similar manner, we define the scattered power as
\begin{eqnarray}
P_s(\omega)=\int_{\partial V}\textbf{S}_s(\textbf{r}',\omega)\cdot\hat{\textbf{n}}d^2r' \ .
\end{eqnarray}
We then obtain from (\ref{div current s}) that
\begin{eqnarray}
\label{Ps}
P_s(\omega)=-\frac{ck(\omega)}{2\pi^2}\int_V\text{Im}(\textbf{E}_s^*(\textbf{r}',\omega)\cdot\textbf{P}(\textbf{r}',\omega))d^3r'\ .
\end{eqnarray}
We define the extinguished power $P_e$ to be the total power lost from the
incident field due to absorption and scattering:
\begin{eqnarray}
P_e(\omega)=P_a(\omega)+P_s(\omega)\  .
\end{eqnarray}
It follows from (\ref{Pa}) and (\ref{Ps}) that the extinguished power is given by
\begin{eqnarray}
\label{extinguished power}
P_e(\omega)=\frac{ck(\omega)}{2\pi^2}\int_V\text{Im}(\textbf{E}_{inc}^*(\textbf{r}',\omega)\cdot\textbf{P}(\textbf{r}',\omega))d^3r' \ .
\end{eqnarray}
{We note that if either Eqs.~(\ref{quadratic polarization}) or (\ref{cubic polarization}) for the polarization is inserted into 
the above expression, we see that the power extinguished from the incident field
is due to interference between the incident field and the total field within the volume of the scatterer.}

We can now rewrite (\ref{extinguished power}) in terms of the scattering amplitude, provided that the incident field is a  plane wave of the form
\begin{eqnarray}
\textbf{E}_{inc}(\textbf{r},\omega)=\textbf{E}_0(\omega) e^{ik(\omega)\hat{\textbf{s}}\cdot\textbf{r}}\ ,
\end{eqnarray}
where $\hat{\textbf{s}}$ is the direction of propagation.
{
Upon comparing (\ref{extinguished power}) and (\ref{scattering amp}), we  obtain
the optical theorem 
\begin{eqnarray}
\label{OPT f}
P_e(\omega)=\frac{c}{2\pi^2k(\omega)}\text{Im}\textbf{A}(\hat{\textbf{s}},\omega)\cdot\textbf{E}^*_0(\omega)\ .
\end{eqnarray}
Using this result, we see
that the time-averaged extinguished power is given by 
\begin{eqnarray}
\label{OPT t}
\bar P_e=\int_{0}^{\infty}P_e(\omega)d\omega=\frac{c}{2\pi^2}\int_{0}^{\infty}\frac{1}{k(\omega)}\text{Im}\textbf{A}(\hat{\textbf{s}},\omega)\cdot\textbf{E}^*_0(\omega)d\omega\ .
\end{eqnarray}}

{
We note that the optical theorem (\ref{OPT f}) applies to both linear and nonlinear media. In Section \ref{applications}, we specialize this result to the cases of quadratic and cubic nonlinearities. Here we remark that in the case of a linear medium with an incident monochromatic field of the form
\begin{equation}
\label{source}
\textbf{E}_{inc}(\textbf{r},\omega)=e^{ik(\omega)\hat{\textbf{s}}\cdot\textbf{r}}(\textbf{E}_0\delta(\omega-\Omega)+\textbf{E}_0^*\delta(\omega+\Omega))\ ,
\end{equation}
(\ref{OPT t})  becomes
\begin{eqnarray}
\bar P_e=\frac{c}{2\pi^2k(\Omega)}\text{Im}\textbf{A}(\hat{\textbf{s}},\Omega)\cdot\textbf{E}_0^*\  .
\end{eqnarray}
We have thus recovered the familiar form of the optical theorem (\ref{opt thm}).
Finally, we note that (\ref{OPT f}) is an \emph{exact} result. That is, it has not been derived by making use of the asymptotic behavior of the electric field.}

\section{Second- and Third-Order Nonlinearities}
\label{applications}
{
Evidently, in order to apply the optical theorem~(\ref{OPT t}) it is necessary to first obtain the scattering amplitude. In this section we calculate the scattering amplitude for second- and third-order nonlinearities. We begin with the case of second-order nonlinearity and, for simplicity, discuss only the problem of second-harmonic generation (SHG). In this setting, we analyze the scattering of an incident monochromatic wave from a {spherical particle whose size is small compared to the wavelength}.  Next, we turn our attention to the case of third-order nonlinearity, where we restrict our attention to the Kerr effect. Once again, we calculate the extinguished power for a small particle and study the associated resonant scattering.}

We note that the method we develop for calculating the scattering of light from a small nonlinear inhomogeneity may be of independent interest. In particular, it is readily extended to collections of small inhomogeneities, which is a physical setting that arises in applications to biomedical imaging and nonlinear microscopy~\cite{Vigoureux,Takahashi,Zayats_1,Zayats_2,Harutyunyan,Palomba_1,Palomba_2}. We plan to report the results of such calculations elsewhere. 

\subsection{Second Order Nonlinearity}

We consider SHG excited by a monochromatic incident field of frequency $\Omega$. We assume that the second-order susceptibility is sufficiently weak so that the condition 
\begin{equation}
\label{weak}
\sum_{\omega_1+\omega_2=\omega}\chi^{(2)}_{ijk}(\textbf{r};\omega_1,\omega_2)E_j(\textbf{r},\omega_1)E_k(\textbf{r},\omega_2)\ll\chi^{(1)}_{ij}(\textbf{r};\omega)E_j(\omega) 
\end{equation}
is obeyed. We then find that the wave equation (\ref{wave eqn}) together with (\ref{quadratic polarization}) and the permutation symmetry of $\chi^{(2)}_{ijk}$
gives rise to the pair of coupled
wave equations
\begin{eqnarray}
\label{E1}
\nonumber
&&\nabla\times\nabla\times\textbf{E}(\textbf{r},\Omega)-k^2(\Omega)\textbf{E}(\textbf{r},\Omega)=4\pi k^2(\Omega)( \chi_{ij}^{(1)}(\textbf{r},\Omega)E_j(\textbf{r},\Omega)+2\chi_{ijk}^{(2)}(\textbf{r},2\Omega,-\Omega)E_j(\textbf{r},2\Omega)E_k^*(\textbf{r},\Omega)) \ , \\ 
\\
\nonumber
&&\nabla\times\nabla\times\textbf{E}(\textbf{r},2\Omega)-k^2(2\Omega)\textbf{E}(\textbf{r},2\Omega)=4\pi k^2(2\Omega)( \chi_{ij}^{(1)}(\textbf{r},2\Omega)E_j(\textbf{r},2\Omega)+\chi_{ijk}^{(2)}(\textbf{r},\Omega,\Omega)E_j(\textbf{r},\Omega)E_k(\textbf{r},\Omega))  \ , \\
\label{E2}
\end{eqnarray}
for the electric fields at the frequencies $\Omega$ and $2\Omega$. Note that we have not accounted for the formation of higher harmonics, consistent with the condition (\ref{weak}).

It follows immediately from (\ref{soln}) that the solutions to (\ref{E1}) and (\ref{E2}) are given by
\begin{eqnarray}
\label{soln1}
E_i(\textbf{r},\Omega)&=&E_{inc,i}(\textbf{r},\Omega)+k^2(\Omega)\int d^3 r' \chi_{jk}^{(1)}(\textbf{r}',\Omega)G_{ij}(\textbf{r},\textbf{r}';\Omega)E_k(\textbf{r}',\Omega)\notag\\
&+& 2k^2(\Omega) \int d^3 r' \chi_{jkl}^{(2)}(\textbf{r}',2\Omega,-\Omega)G_{ij}(\textbf{r},\textbf{r}';\Omega)E_k(\textbf{r}',2\Omega)E_l^*(\textbf{r}',\Omega) \ ,\\
E_i(\textbf{r},2\Omega)&=&k^2(2\Omega)\int d^3 r' \chi_{jk}^{(1)}(\textbf{r}',2\Omega)G_{ij}(\textbf{r},\textbf{r}';2\Omega)E_k(\textbf{r}',2\Omega) \notag \\
&+& k^2(2\Omega) \int d^3 r' \chi_{jkl}^{(2)}(\textbf{r}',\Omega,\Omega)G_{ij}(\textbf{r},\textbf{r}';2\Omega)E_k(\textbf{r}',\Omega)E_l(\textbf{r}',\Omega) \  .
\label{soln2}
\end{eqnarray}

Suppose that the scattering medium is a small ball of radius $a$ with $k(\Omega)a \ll 1$ and 
$k(2\Omega)a \ll 1$. The susceptibilities are taken to be $\chi^{(1)}_{ij}(\textbf{r};\omega)=\eta_{ij}^{(1)}$ and $\chi^{(2)}_{ijk}(\textbf{r};\omega)=\eta^{(2)}_{ijk}$ for $|\textbf{r}|\le a$ and to vanish for $|\textbf{r}|>a$.
Eqs.~(\ref{soln1}) and (\ref{soln2}) thus become
\begin{eqnarray}
\label{SHG int1}
E_i(\textbf{r},\Omega)&=&E_{inc,i}(\textbf{r},\Omega)+k^2(\Omega)\eta_{jk}^{(1)}\int_{|\textbf{r}'|\le a} d^3 r'G_{ij}(\textbf{r},\textbf{r}';\Omega)E_k(\textbf{r}',\Omega)\notag\\
&+&2k^2(\Omega)\eta_{jkl}^{(2)}\int_{|\textbf{r}'|\le a} d^3 r' G_{ij}(\textbf{r},\textbf{r}';\Omega)E_k(\textbf{r}',2\Omega)E_l^*(\textbf{r}',\Omega) \ , \\
E_i(\textbf{r},2\Omega)&=&k^2(2\Omega)\eta_{jk}^{(1)}\int_{|\textbf{r}'|\le a} d^3r' G_{ij}(\textbf{r},\textbf{r}';2\Omega)E_k(\textbf{r}',2\Omega)\notag\\
&+&k^2(2\Omega)\eta_{jkl}^{(2)}\int_{|\textbf{r}'|\le a} d^3 r'G_{ij}(\textbf{r},\textbf{r}';2\Omega)E_k(\textbf{r}',\Omega)E_l(\textbf{r}',\Omega) \ .
\label{SHG int2}
\end{eqnarray}
Using the asymptotic form of the Green's function given in (\ref{G asymp}), we find that the scattered fields are of the form
\begin{eqnarray}
E_i^s (\textbf{r},\Omega)&=& A_i(\textbf{r},\Omega) \frac{e^{ik(\Omega)r}}{r} \ , \\
E_i(\textbf{r},2\Omega)&=&A_i(\textbf{r},2\Omega)\frac{e^{ik(2\Omega)r}}{r} \ ,
\end{eqnarray}
where the scattering amplitudes are defined by
\begin{eqnarray}
\label{SHG amp}
A_i(\textbf{r},\Omega)&=&\frac{4\pi}{3}a^3(\delta_{ij}-\hat{r}_i\hat{r}_j)k^2(\Omega)(\eta_{jk}^{(1)}E_k(\textbf{0},\Omega)+2\eta_{jkl}^{(2)}E_k(\textbf{0},2\Omega)E_l^*(\textbf{0},\Omega)) \ , \\
A_i(\textbf{r},2\Omega)&=&\frac{4\pi}{3}a^3(\delta_{ij}-\hat{r}_i\hat{r}_j)k^2(2\Omega)(\eta_{jk}^{(1)}E_j(\textbf{0},2\Omega)+\eta_{jkl}^{(2)}E_k(\textbf{0},\Omega)E_l(\textbf{0},\Omega))\ .
\end{eqnarray}
Here we have used the fact that the radius of the scatterer is small, which leads to the identity
\begin{eqnarray}
\int_{|\textbf{r}'|\le a} d^3 r' e^{-ik(\omega)\hat{\textbf{r}}\cdot\textbf{r}'}g(k(\omega)\textbf{r}')=\frac{4\pi}{3}a^3g(\textbf{0})(1+O(k(\omega)a))\ ,
\end{eqnarray}
for some function $g$.
{
The local fields $E_i(\textbf{0},\Omega)$ and $E_i(\textbf{0},2\Omega)$ can be calculated perturbatively as shown in Appendix~\ref{local_fields_quadratic}}. 

\subsection{Third Order Nonlinearity}

The treatment of the Kerr effect parallels that of SHG. We consider the Kerr effect excited by a monochromatic incident field of frequency $\Omega$. We assume that the third-order susceptibility is sufficiently weak so that the condition 
\begin{equation}
\label{weak again}
\sum_{\omega_1+\omega_2+\omega_3=\omega}\chi^{(3)}_{ijkl}(\textbf{r};\omega_1,\omega_2)E_j(\textbf{r},\omega_1)E_k(\textbf{r},\omega_2)E_l(\textbf{r},\omega_3) \ll\chi^{(1)}_{ij}(\textbf{r};\omega)E_j(\omega)
\end{equation}
is obeyed. We then find that (\ref{wave eqn}) together with (\ref{cubic polarization}) and the permutation symmetry of $\chi^{(3)}_{ijkl}$ gives rise to the wave equation
\begin{eqnarray}
\label{E3}
\nabla\times\nabla\times\textbf{E}(\textbf{r},\Omega)-k^2(\Omega)\textbf{E}(\textbf{r},\Omega)=&&4\pi k^2(\Omega)( \chi_{ij}^{(1)}(\textbf{r},\Omega)E_j(\textbf{r},\Omega)\notag\\
&&+3\chi_{ijkl}^{(3)}(\textbf{r},\Omega,\Omega,-\Omega)E_j(\textbf{r},\Omega)E_k(\textbf{r},\Omega)E_l^*(\textbf{r},\Omega)) \ . 
\end{eqnarray}
Note that we have not accounted for the formation of higher harmonics, consistent with the condition (\ref{weak again}).
It follows immediately from (\ref{soln}) that the solution to (\ref{E3}) is given by
\begin{eqnarray}
\label{soln3}
E_i(\textbf{r},\Omega)&=&E_{inc,i}(\textbf{r},\Omega)+k^2(\Omega)\int d^3 r' \chi_{jk}^{(1)}(\textbf{r}',\Omega)G_{ij}(\textbf{r},\textbf{r}';\Omega)E_k(\textbf{r}',\Omega)\notag\\
&+& 3k^2(\Omega) \int d^3 r' \chi_{jklm}^{(3)}(\textbf{r}',\Omega,\Omega,-\Omega)G_{ij}(\textbf{r},\textbf{r}';\Omega)E_k(\textbf{r}',\Omega)E_l(\textbf{r}',\Omega)E_m^*(\textbf{r}',\Omega)) \ ,
\end{eqnarray}

Suppose that the scattering medium is a small ball of radius $a$ with $k(\Omega)a \ll 1$. The susceptibilities are taken to be $\chi^{(1)}_{ij}(\textbf{r};\omega)=\eta_{ij}^{(1)}$ and $\chi^{(3)}_{ijkl}(\textbf{r};\omega)=\eta^{(3)}_{ijkl}$ for $|\textbf{r}|\le a$ and to vanish for $|\textbf{r}|>a$.
Eq.~(\ref{soln3}) thus become
\begin{eqnarray}
\label{Kerr int}
E_i(\textbf{r},\Omega)&=&E_{inc,i}(\textbf{r},\Omega)+k^2(\Omega)\eta_{jk}^{(1)}\int_{|\textbf{r}'|\le a} d^3 r'G_{ij}(\textbf{r},\textbf{r}';\Omega)E_k(\textbf{r}',\Omega)\notag\\
&+&3k^2(\Omega) \eta^{(3)}_{jklm}\int_{|\textbf{r}'|\le a}  d^3 r' G_{ij}(\textbf{r},\textbf{r}';\Omega)E_k(\textbf{r}',\Omega)E_l(\textbf{r}',\Omega)E_m^*(\textbf{r}',\Omega))\ .
\end{eqnarray}
Using the asymptotic form of the Green's function given in (\ref{G asymp}), we find that the scattered field is of the form
\begin{eqnarray}
E_i^s (\textbf{r},\Omega)&=& A_i(\textbf{r},\Omega) \frac{e^{ik(\Omega)r}}{r} \ . 
\end{eqnarray}
where the scattering amplitude is defined by
\begin{eqnarray}
A_i(\textbf{r},\Omega)&=&\frac{4\pi}{3}a^3(\delta_{ij}-\hat{r}_i\hat{r}_j)k^2(\Omega)(\eta_{jk}^{(1)}E_k(\textbf{0},\Omega)+3\eta_{jklm}^{(3)}E_k(\textbf{0},\Omega)E_l(\textbf{0},\Omega)) E_m^*(\textbf{0},\Omega))\ , 
\end{eqnarray}
{
Once again, we calculate the local fields $E_i(\textbf{0},\Omega)$ perturbatively,
as shown in Appendix~\ref{local_fields_cubic}.} 

\section{Numerical Results}

In this section we apply the optical theorem (\ref{OPT t}) to linear, second- and third-order nonlinear media. We present numerical results for several cases of interest, including second harmonic generation and the Kerr effect. {We will see that the effect of the nonlinearities is to modify the linear scattering resonance of small scatterers.}

\subsection{Linear Response}

We consider an isotropic medium with $\eta_{ij}^{(1)}=\eta^{(1)}\delta_{ij}$ and assume that all the higher-order susceptibilities vanish.
The incident field is taken to be a unit-amplitude plane wave of the form $\textbf{E}_{inc}= \textbf{E}_0 \exp(ik_0\hat{\textbf{s}}\cdot \textbf{r})$ with $ \textbf{E}_0=E_0\hat{\textbf{x}}$ and $\hat{\textbf{s}} = \hat{\textbf{z}}$. Using (98) and (91), the extinguished power becomes
\begin{equation}
\label{linear_Pe}
P_e=  \frac{8\Omega}{3}E_0^2 a^3\text{Im}\left(\frac{\eta^{(1)}}{1-\frac{4\pi}{3}k^2a^3 \eta^{(1)}G_R }\right) \ .
\end{equation}
We can write the above formula in a more familiar form in terms of the renormalized polarizability
$\alpha$, which is defined as
\begin{equation}
\alpha = \frac{\alpha_0}{1-k^2\alpha_0\left[1/a + i \frac{2}{3}k\right]} \ ,
\end{equation}
where $\alpha_0$ is the zero-frequency polarizability, which is defined in terms of the linear dielectric
permittivity $\epsilon^{(1)}$. Here
\begin{equation}
\alpha_0 = \frac{{\epsilon^{(1)}}-1}{{\epsilon^{(1)}}+2}a^3\  ,
\end{equation}
where $\epsilon^{(1)}=1+4\pi\eta^{(1)}$. We find that (\ref{linear_Pe}) becomes
\begin{equation}
P_e=  \frac{2\Omega}{\pi}E_0^2 \text{Im}\alpha \ .
\end{equation}
In Figure~1 we illustrate the frequency dependence of the extinguished power for a  dielectric scatterer of size $a=100$nm with $\epsilon^{(1)}=-5.28$. We see that there is a scattering resonance at $ka\approx 0.7$.
\begin{figure}
\includegraphics[width=0.8\linewidth]{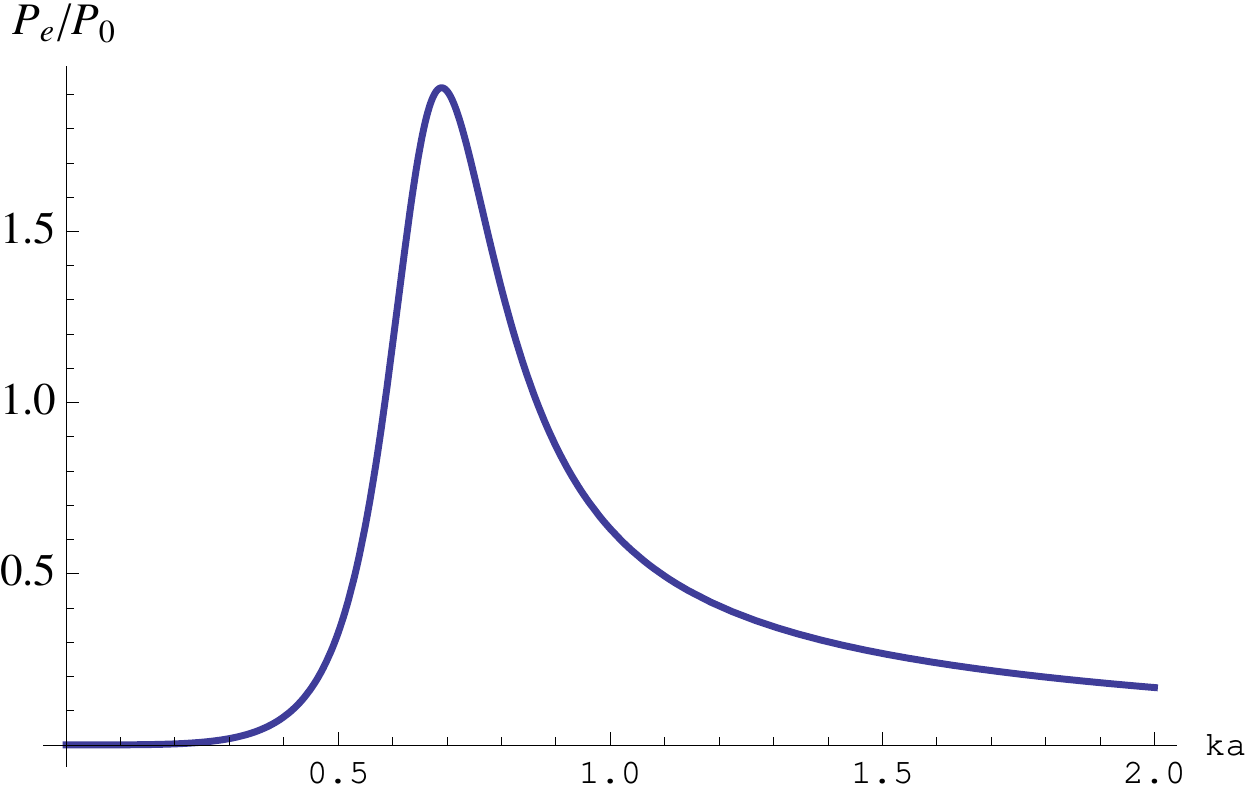}
\caption{Frequency dependence of extinguished power for a single linear scatterer.
{Here $P_0=a^2cE_0^2$. }}
\label{Linear 1pt}
\end{figure}

\subsection{Second Harmonic Generation}

{
We consider the case of a medium with isotropic $\eta^{(1)}$ and $\eta^{(2)}$ obeying permutation symmetry. That is, $\eta_{ij}^{(1)}=\eta^{(1)}\delta_{ij}$ and
$\eta^{(2)}_{11}=\eta^{(2)}$,
with the other $\eta^{(2)}_{ijk}$ vanishing.
We also assume that the incident field $\textbf{E}_{inc}$ points in the $x$-direction and the direction of observation $\hat{\textbf{s}}$ is taken to be in the $z$-direction. It follows from (\ref{Pe_SHG}) that the extinguished power $P_e$ is given by 
\begin{eqnarray}
P_e=  \frac{8\Omega}{3} a^3\text{Im}\left((\eta^{(1)}E_1^{(0)}(\textbf{0},\Omega)+\eta^{(1)}E_1^{(2)}(\textbf{0},\Omega)+2\eta^{(2)}_{111}(E_1^{(0)}(\textbf{0},\Omega))^*E_1^{(1)}(\textbf{0},2\Omega))E_{inc,1}^*(\textbf{0},\Omega)\right)\ .\notag\\
\end{eqnarray}}

In Figure~2 we illustrate the frequency dependence of the extinguished power for SHG. The scatterer size is $a=100$nm and $E_0$ is taken to have unit amplitude. Plots are shown for $\epsilon=\eta^{(2)}E_0/\eta^{(1)}= 0, 0.1, 0.2, 0.3$. {We see that the resonance shifts to lower frequencies and its amplitude increases by more than a factor of two relative to the linear case. The physical situation considered corresponds to the material $\beta$-${\rm Ba B_2 O_4}$~\cite{Boyd}. {We note that there does not appear to be a  simple physical argument to predict the extent or direction of the reported frequency shifts.}

\begin{figure}
\includegraphics[width=0.8\linewidth]{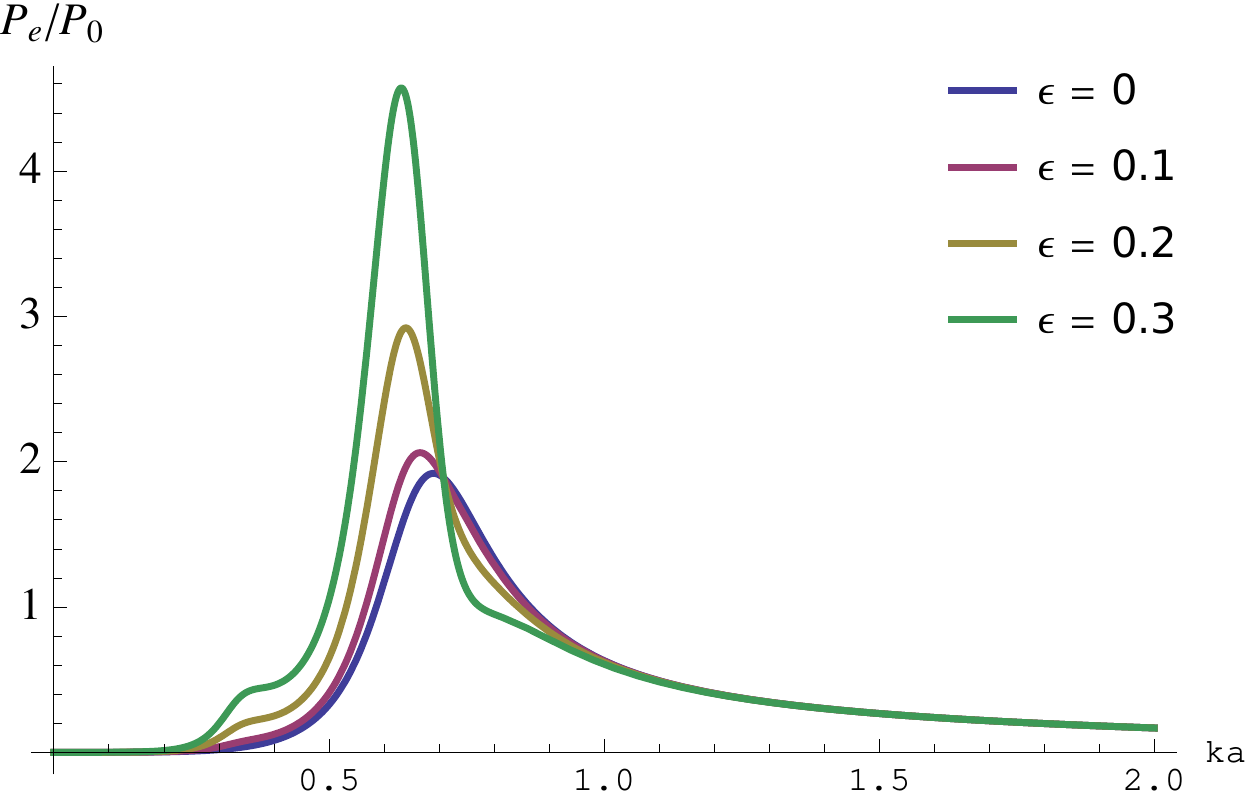}
\caption{(Color online) Frequency dependence of extinguished power for a single nonlinear scatterer with SHG. {Here $P_0=a^2cE_0^2$. }}
\label{SHG 1pt}
\end{figure}

\subsection{Kerr Effect}

{We consider the case of a medium with isotropic $\eta^{(1)}$ and $\eta^{(3)}$ obeying the permutation symmetry. That is, $\eta_{ij}^{(1)}=\eta^{(1)}\delta_{ij}$ and $\eta_{1111}^{(3)}=\eta^{(3)}$, with all other $\eta_{ijkl}^{(3)}$ vanishing. We also assume that the incident field $\textbf{E}_{inc}$ points in the $x$-direction and the direction of observation $\hat{\textbf{s}}$ is taken to be in the $z$-direction. It follows from (\ref{Pe_kerr}) that the extinguished power $P_e$ is given by 
\begin{eqnarray}
P_e=  \frac{8\Omega}{3} a^3\text{Im}\left((\eta^{(1)}E_1^{(0)}(\textbf{0},\Omega)+\eta^{(1)}E_1^{(1)}(\textbf{0},\Omega)+3\eta^{(3)}_{1111}(E^*)_1^{(0)}(\textbf{0},\Omega)(E)_1^{(0)}(\textbf{0},\Omega)(E)_1^{(0)}(\textbf{0},\Omega))E_{inc,i}^*(\textbf{0},\Omega\right)\ .\notag\\
\end{eqnarray}}

In Figure~3 we illustrate the frequency dependence of the extinguished power for the Kerr effect. The scatterer size is $a=100$nm and $E_0$ is taken to have unit amplitude. Plots are shown for $\epsilon=\eta^{(3)}E_0^2/\eta^{(1)}= 0, 0.01, 0.02$. 
We see that the resonance shifts to higher frequencies relative to the linear case. As may be expected, the effect is less pronounced than in the case of SHG.    {As above, we do not know of a simple physical argument to predict the extent or direction of the reported frequency shifts.}

\begin{figure}
\includegraphics[width=0.8\linewidth]{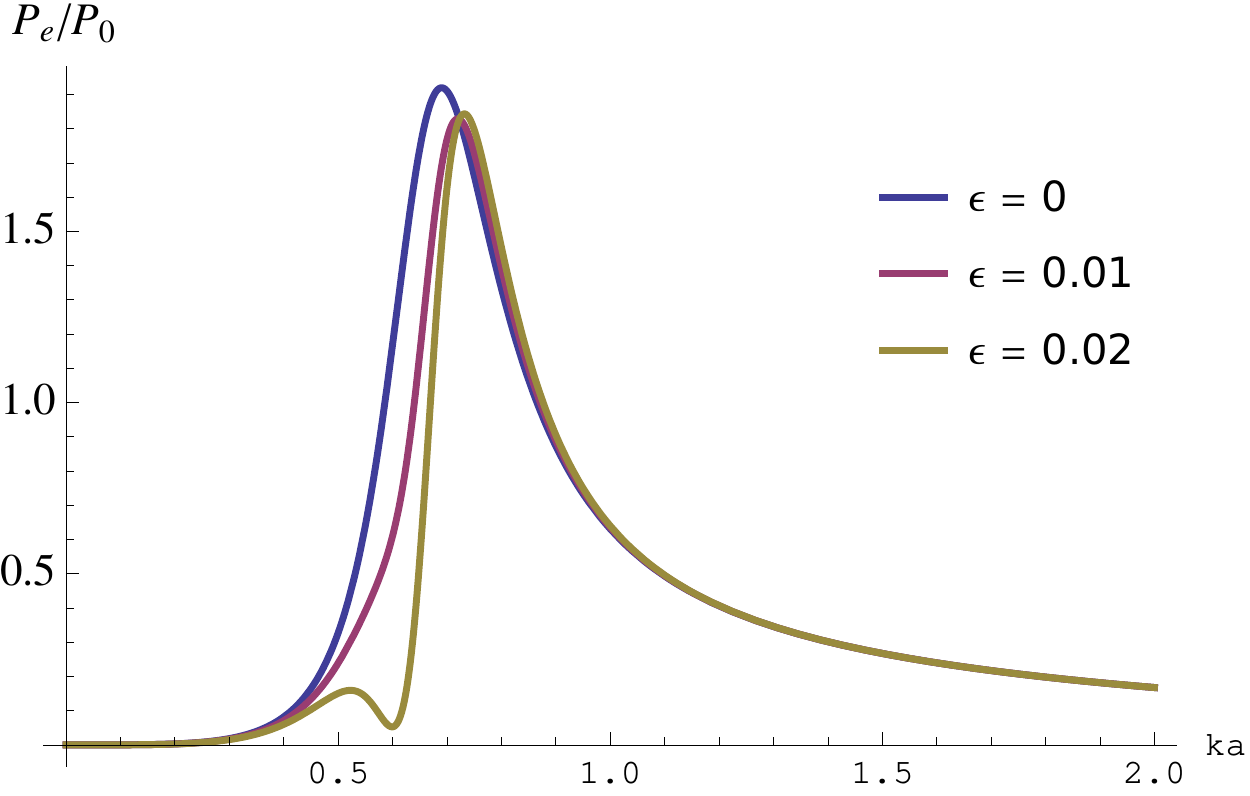}
\caption{(Color online) Frequency dependence of extinguished power for a single  scatterer with Kerr nonlinearity. {Here $P_0=a^2cE_0^2$. }}
\label{Kerr 1pt}
\end{figure}

\section{Application To Near-Field Microscopy}
\label{ch:Application}

Near-field scanning optical microscopy (NSOM) is a widely used experimental tool to overcome
the diffraction limit of optical microscopy~\cite{Novotny}.  In a typical experiment, an apertured probe (often a metallic coated optical fiber) is brought into the near-field of a sample and employed as an optical source. The image is formed by scanning the probe and recording the intensity of light scattered into the far-field. In a reciprocal arrangement, the probe may be used as a detector with the illumination incident from the far-field. In either case, the resolution of the resulting image is controlled by the size of the probe rather than the wavelength of light. 

Apertureless NSOM is an alternative to the above approach in which the illumination and detection both take place in the far-field~\cite{Novotny,Keilmann}. The experimental setup is illustrated in Figure~\ref{NSOM}, in which an incident field illuminates a metallic tip that is placed in the near-field of the sample.  The image is obtained by scanning the tip and measuring the scattered field with a detector that is placed in the far-field of the sample and the tip. 

A refinement of apertureless NSOM is to introduce a fluorescent tip, which allows for the spectral isolation of the detected light and improvement in SNR by background suppression~\cite{Lewis}. Spectral isolation may also be achieved by utilizing a tip that has a nonlinear optical response~\cite{Vigoureux,Takahashi,Zayats_1,Zayats_2,Harutyunyan,Palomba_1,Palomba_2}. This approach has the advantage that it is not affected by fluorescent photobleaching. Experiments in which SHG, THG and four-wave mixing have been utilized for aptertureless NSOM in a dark-field configuration have recently been reported ~\cite{Harutyunyan}.

In this section we develop a model for SHG and THG apertureless NSOM. We consider a system in which a nonlinear metallic tip is placed in the near-field of a pair of small dielectric particles, which are taken to have only a linear optical response. The setup is shown in Figure~\ref{NSOM} and the mathematical details are presented in Appendix C. We begin with the case of SHG. The sample consists of a pair of dielectric spheres of radius $\lambda/(10\pi)$ and susceptibility $\hat{\eta}^{(1)}_{ij}=0.4\delta_{ij}$,  located at the positions $\textbf{r}_1=(0,\frac{3\lambda}{20\pi},0)$ and $\textbf{r}_2=(0,-\frac{3\lambda}{20\pi},0)$, which corresponds to a separation of $\approx \lambda/10$. The tip has radius $a=\lambda/(10\pi)$, linear susceptibility $\eta^{(1)} =- 0.4$, second order susceptibility
$\eta^{(2)}_{111}$ with $\eta^{(2)}_{111}E_0/\hat{\eta}^{(1)}=0.2$ and all other
$\eta^{(2)}_{ijkl}$ vanishing, and is scanned in the planes $x=x_0$. In all numerical experiments, the incident electric field is a plane wave polarized in the $x$-direction with wave vector $k(\Omega)\hat{\textbf z}$. In Figure~\ref{SHGPe} images of
the extinguished power are shown in three different scan planes corresponding to $x_0=2a$, $2.5a$ and $3a$. We see that the scatterers are well resolved with subwavelength separation  in the closest scan plane and that the resolution degrades rapidly with distance from the plane $x=0$. Qualitatively similar results are found for the intensity of scattered second harmonic light, as illustrated in Figure~\ref{SHGA}.

\begin{figure}
\vspace{-1in}
\includegraphics[width=0.9\linewidth]{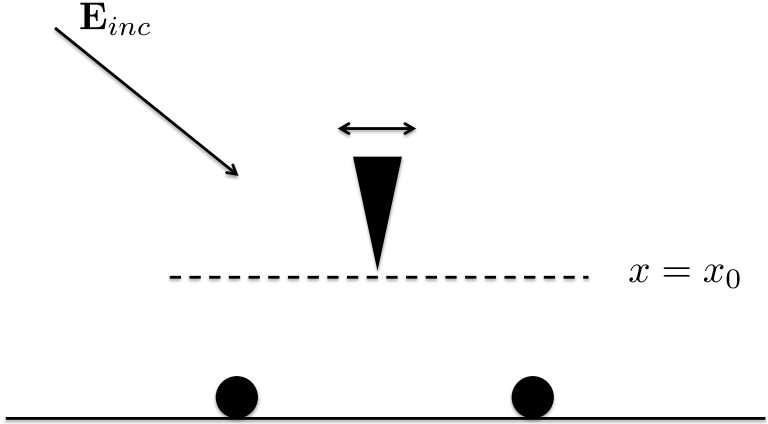}
\vspace{-1in}
\caption{Illustrating the apertureless NSOM experiment.}
\label{NSOM}
\end{figure}

Next we consider the case of THG. The setup is the same as in the case of SHG, except that the third order susceptibility of the tip is $\eta^{(3)}_{1111}$, with $\eta^{(3)}_{1111}E_0^2/\hat{\eta}^{(1)}=0.2$ and all other $\eta^{(3)}_{ijkl}$ vanishing. Once again, we find that the scatterers are better resolved in the closest scan plane. We also note that the relative extinguished power is smaller than in the case of SHG and that the intensity of the scattered third harmonic is correspondingly greater. {See Figures~\ref{THGPe} and \ref{THGA}. In Figure~\ref{PeComp} the extinguished power along a line in the closest scan plan is compared for SHG and THG. It is found that the the extinguished power for the case of a SHG tip is an {order of magnitude larger} than for a tip exhibiting THG. The corresponding result for the scattering amplitude is shown in Figure~\ref{AComp}.}

\section{Discussion}

In this paper we have presented a generalization of the optical theorem for the scattering of nonlinear electromagnetic waves. We consider in some detail the most important examples of quadratic and cubic nonlinearities. The theory is illustrated for the case of small inhomogeneities in the settings of second harmonic generation and the Kerr effect. In particular, we describe the manner in which scattering resonances are modified by the presence of nonlinearity. As a second application, we consider the problem of computing the signal in a nonlinear near-field microscopy experiment. We note that the use of a tip with a nonlinear optical response affords the possibility of background suppression and spectral isolation of the detected signal. In future work, we plan to study the inverse problem for nonlinear near-field microscopy, whose goal is to reconstruct the linear optical properties of a sample illuminated by a nonlinear tip. This will necessitate the development of a scattering theory that incorporates contributions from the tip and the sample and their respective interactions, as was done for the corresponding linear problem~\cite{Sun}.

\begin{figure}[]
\centering
\subfigure[\  $x_0=\lambda/(5\pi)$]{%
\includegraphics[width=0.3\linewidth]{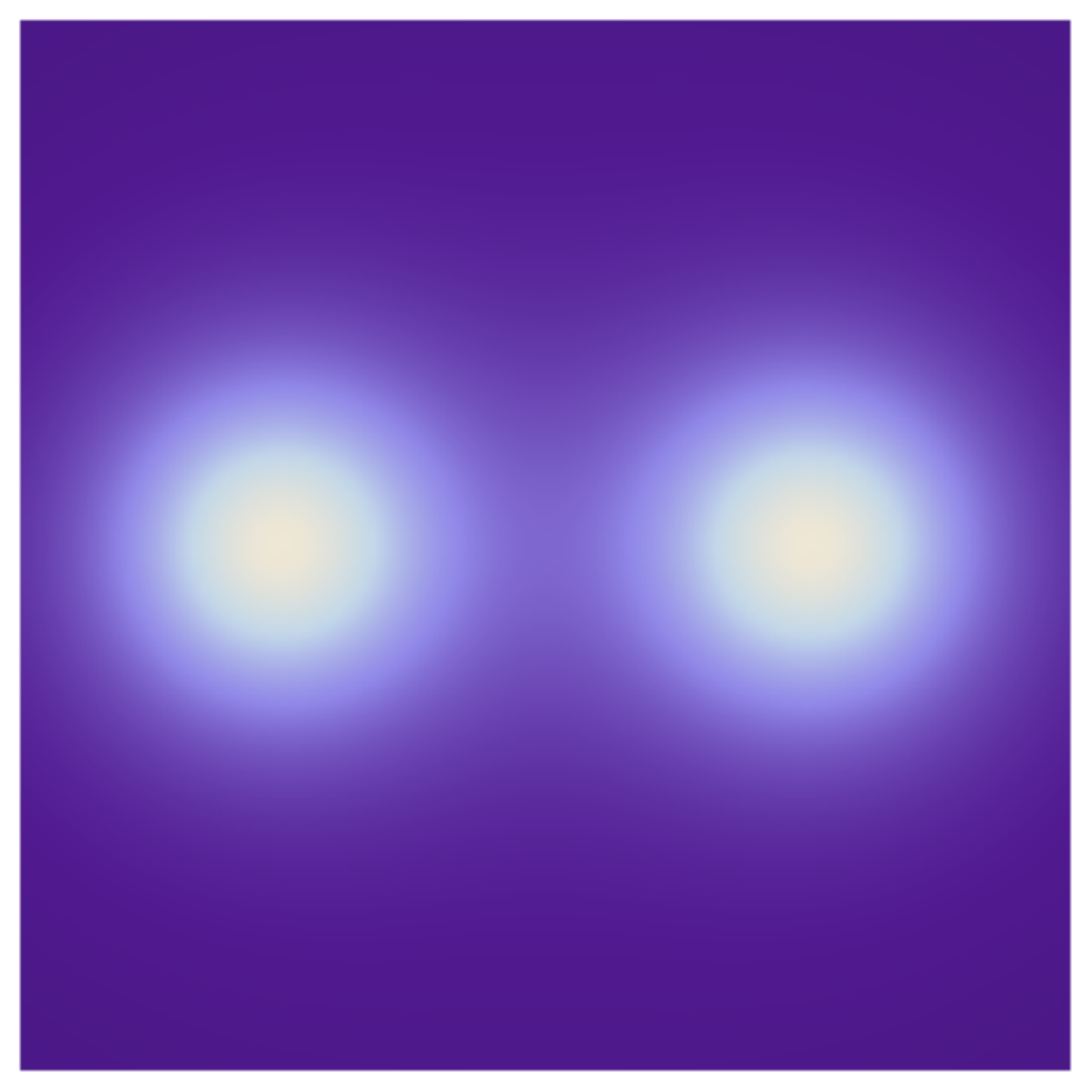}
\label{SHGPe2}}
\subfigure[\ $x_0=\lambda/(4\pi)$]{%
\includegraphics[width=0.3\linewidth]{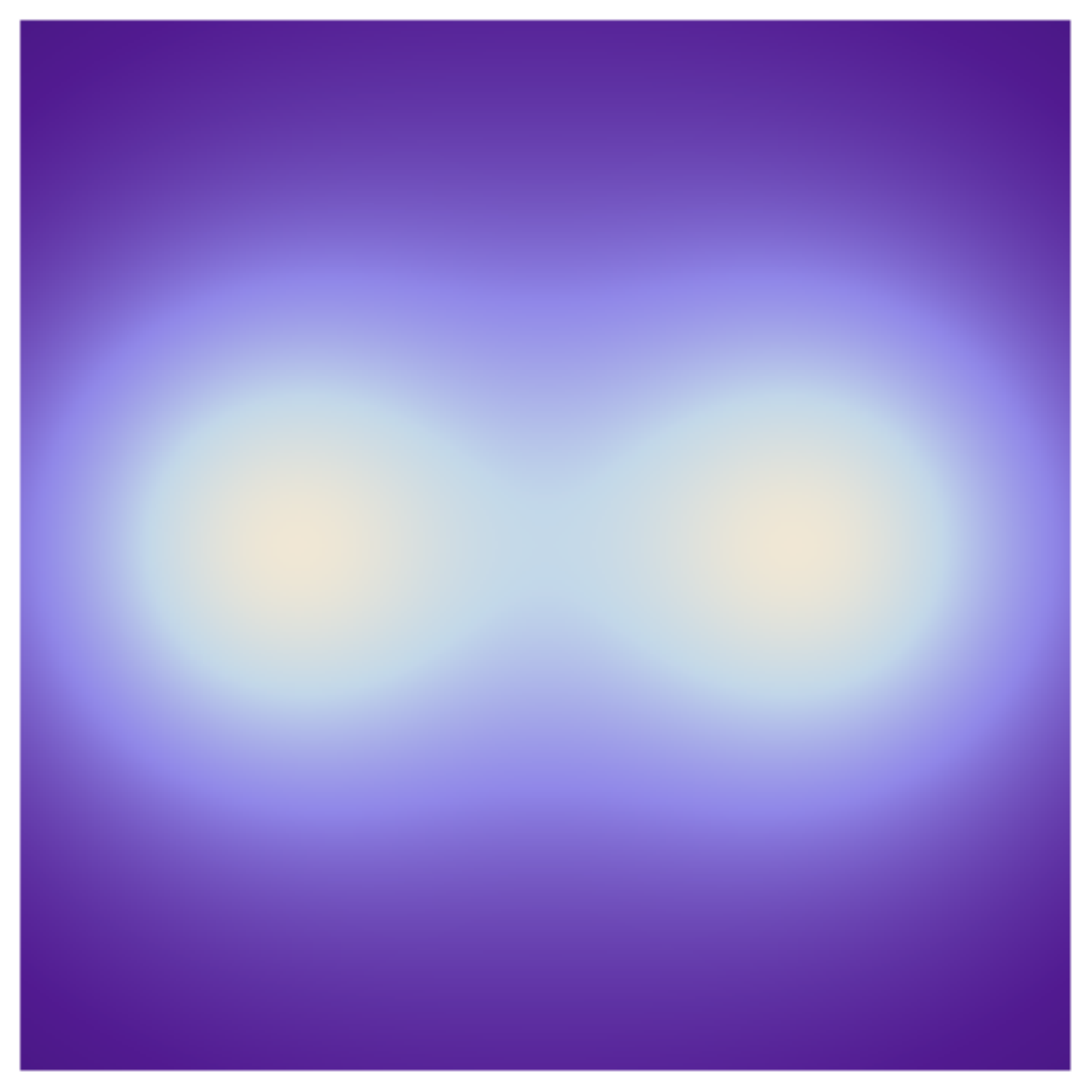}
\label{SHGPe2.5}}
\subfigure[\  $x_0=3\lambda/(10\pi)$]{%
\includegraphics[width=0.3\linewidth]{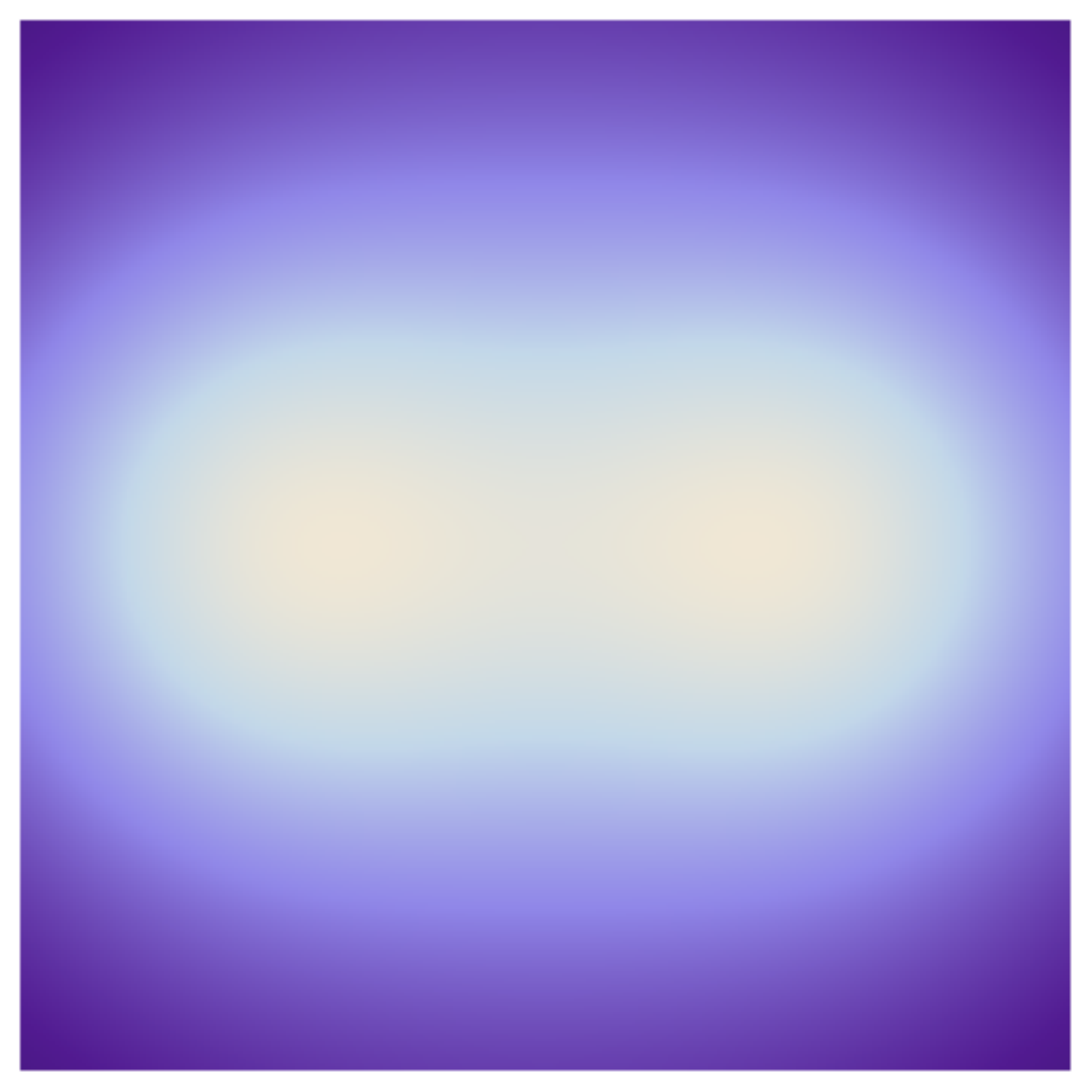}
\label{SHGPe3}}
\caption{(Color online) Apertureless NSOM images of the extinguished power $P_e$ for SHG.
The susceptibility of the tip is $\eta^{(2)}_{111}E_0/\hat{\eta}^{(1)}=0.2$. 
Images are shown in the planes $x=x_0$ as indicated.
The field of view of each image is
$3\lambda/(5\pi) \times 3\lambda/(5\pi)$. }
\label{SHGPe}
\end{figure}

\begin{figure}[]
\centering
\subfigure[\ $x_0=\lambda/(5\pi)$]{%
\includegraphics[width=0.3\linewidth]{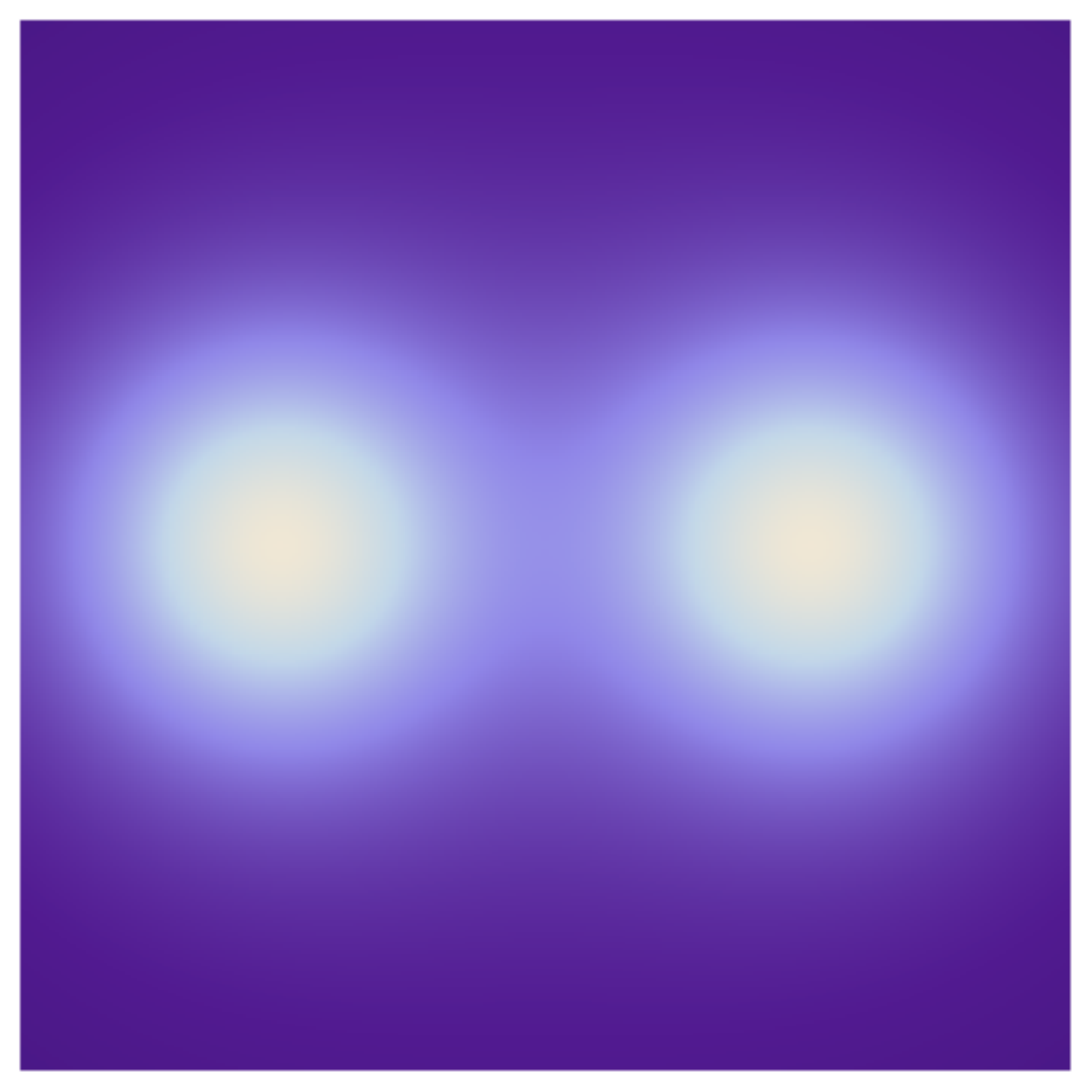}
\label{THGPe2}}
\subfigure[\ $x_0=\lambda/(4\pi)$]{%
\includegraphics[width=0.3\linewidth]{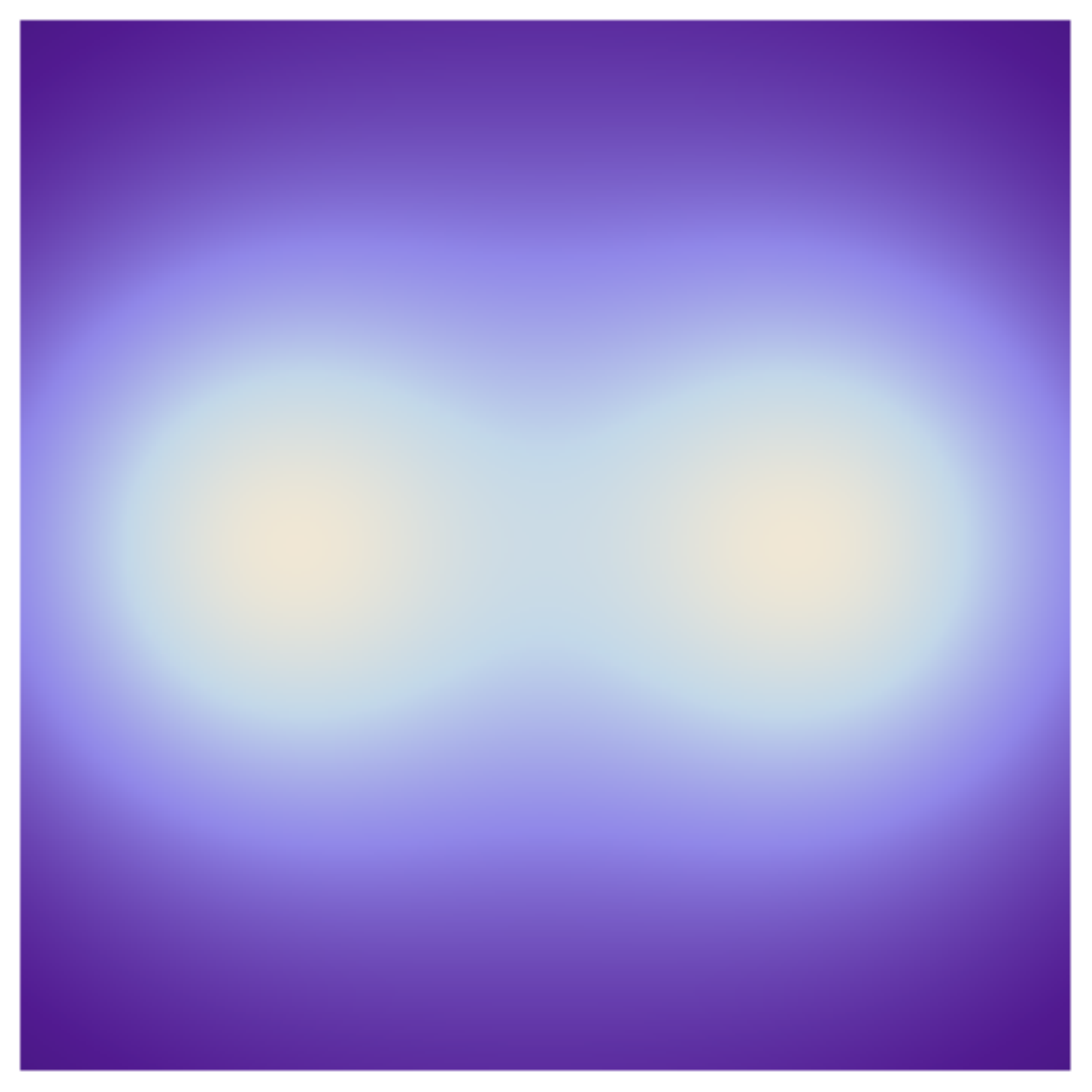}
\label{THGPe25}}
\subfigure[\ $x_0=3\lambda/(10\pi)$]{%
\includegraphics[width=0.3\linewidth]{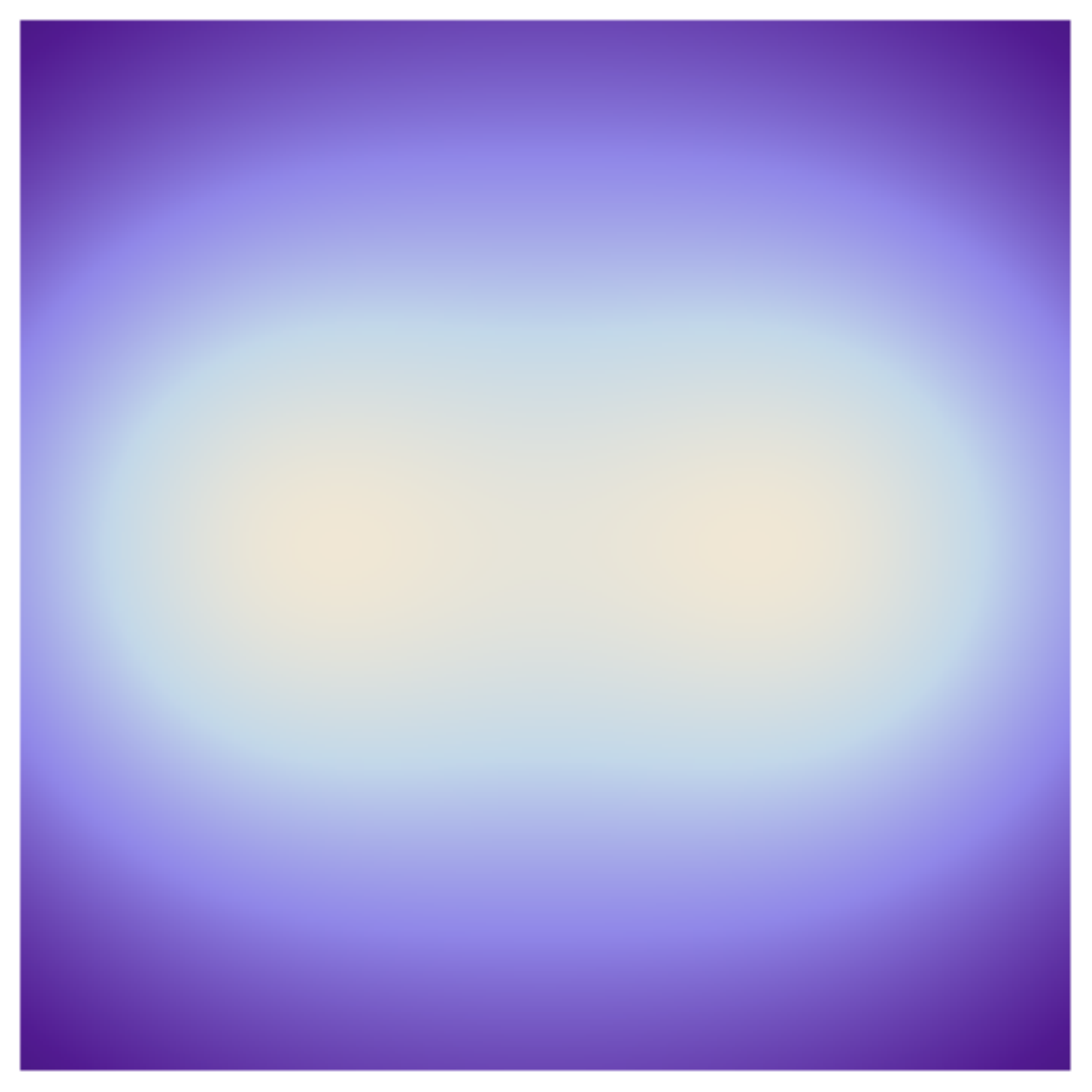}
\label{THGPe3}}
\caption{(Color online) Apertureless NSOM images of the extinguished power $P_e$ for THG.
The susceptibility of the tip is $\eta^{(3)}_{1111}E_0/\hat{\eta}^{(1)}=0.2$. 
Images are shown in the planes $x=x_0$ as indicated.
The field of view of each image is
$3\lambda/(5\pi) \times 3\lambda/(5\pi)$. }
\label{THGPe}
\end{figure}

\begin{figure}[]
\centering
\subfigure[\ $x_0=\lambda/(5\pi)$]{%
\includegraphics[width=0.3\linewidth]{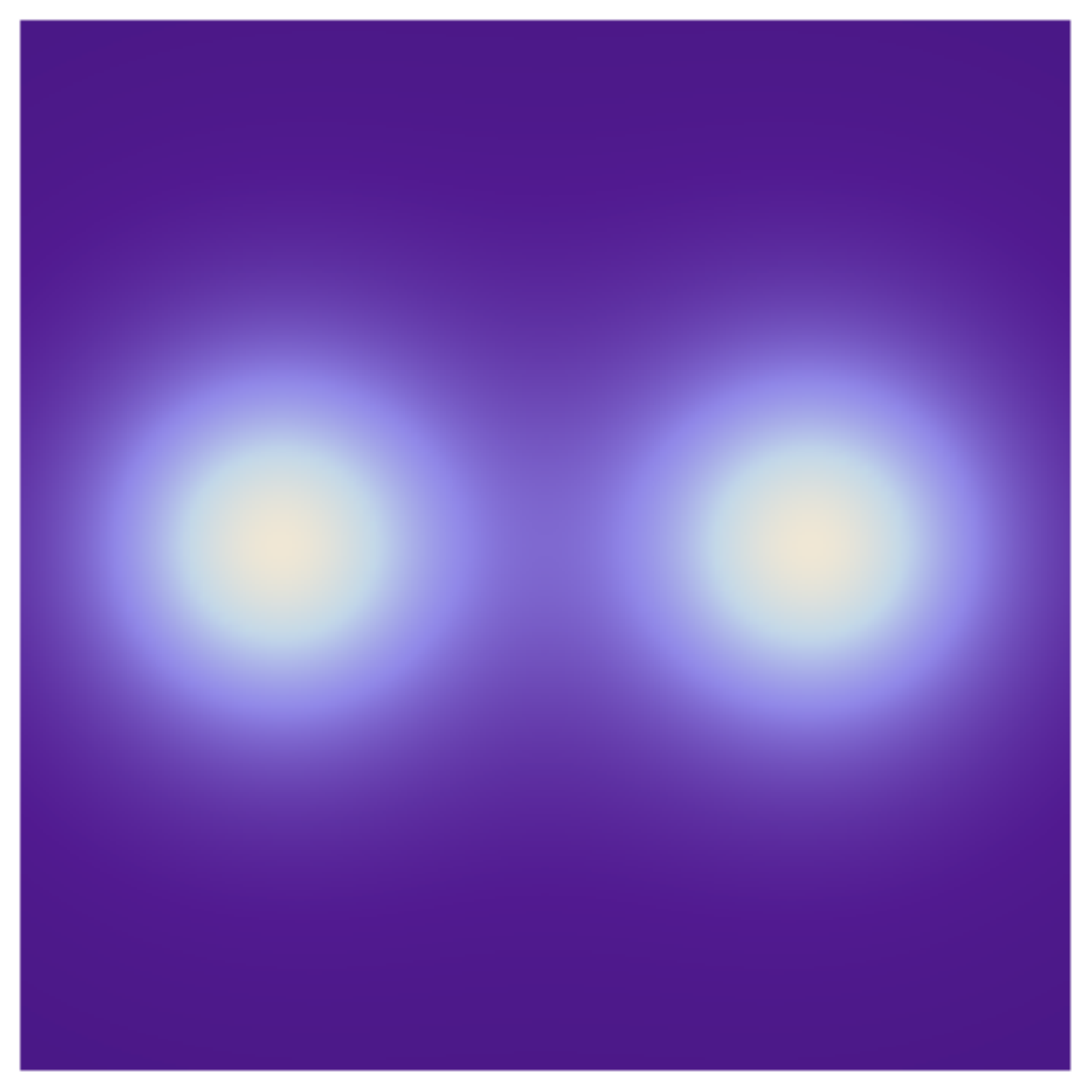}
\label{SHGA2}}
\subfigure[\ $x_0=\lambda/(4\pi)$]{%
\includegraphics[width=0.3\linewidth]{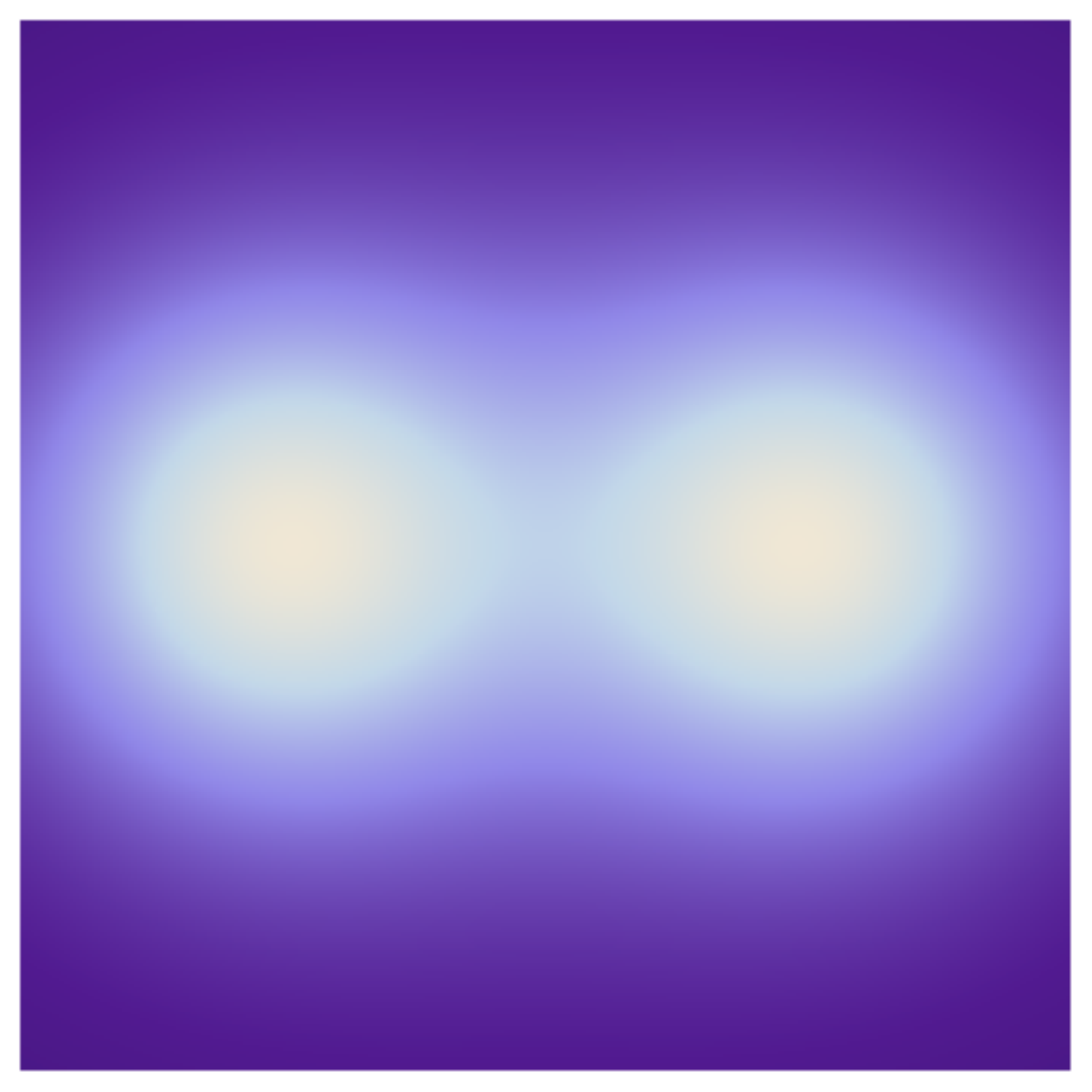}
\label{SHGA2.5}}
\subfigure[\ $x_0=3\lambda/(10\pi)$]{%
\includegraphics[width=0.3\linewidth]{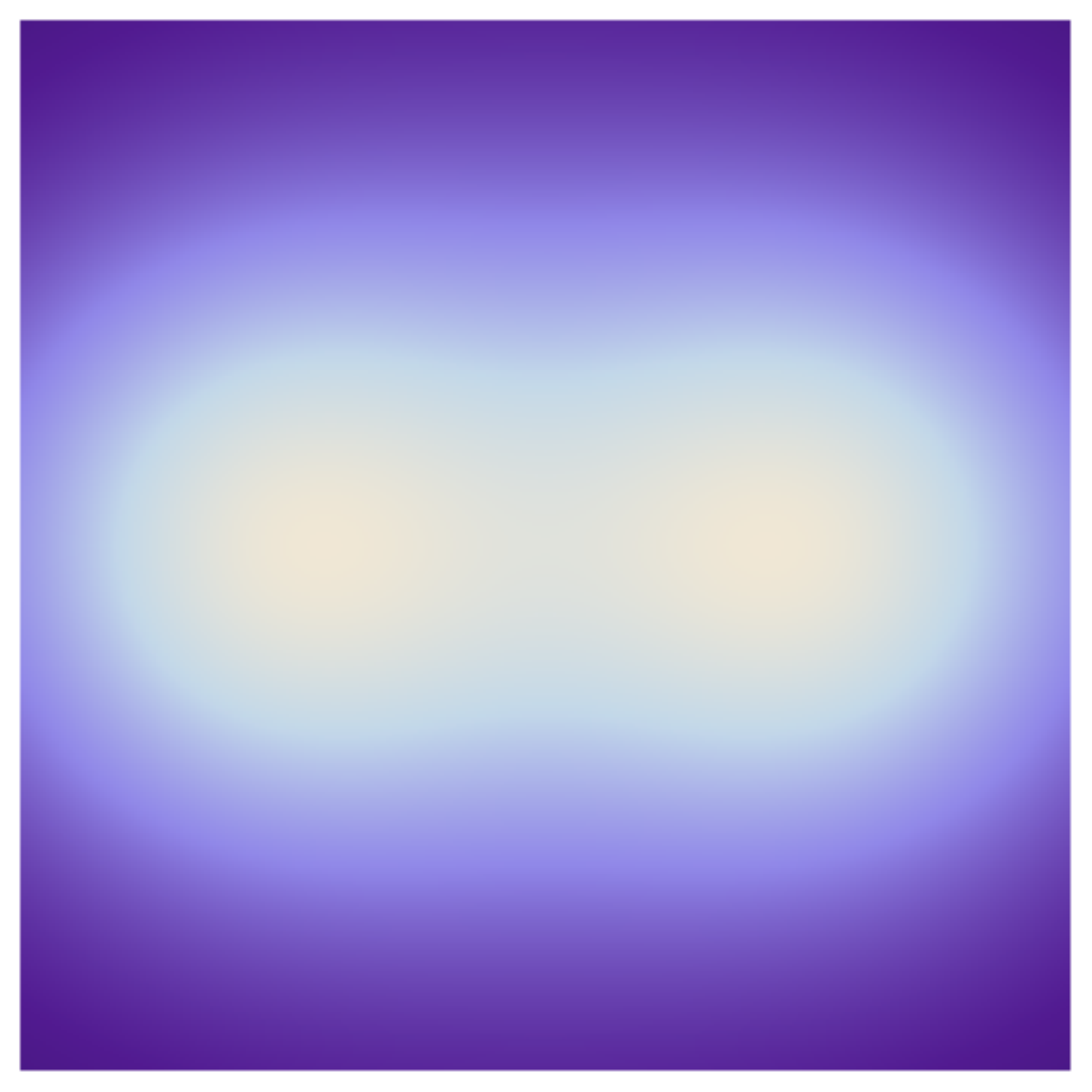}
\label{SHGA3}}
\caption{(Color online) Apertureless NSOM images of the far-field intensity at frequency $2\Omega$ in SHG.
The susceptibility of the tip is $\eta^{(2)}_{111}E_0/\hat{\eta}^{(1)}=0.2$. 
Images are shown in the planes $x=x_0$ as indicated.
The field of view of each image is
$3\lambda/(5\pi) \times 3\lambda/(5\pi)$. }
\label{SHGA}
\end{figure}

\begin{figure}[]
\centering
\subfigure[\ $x_0=\lambda/(5\pi)$]{%
\includegraphics[width=0.3\linewidth]{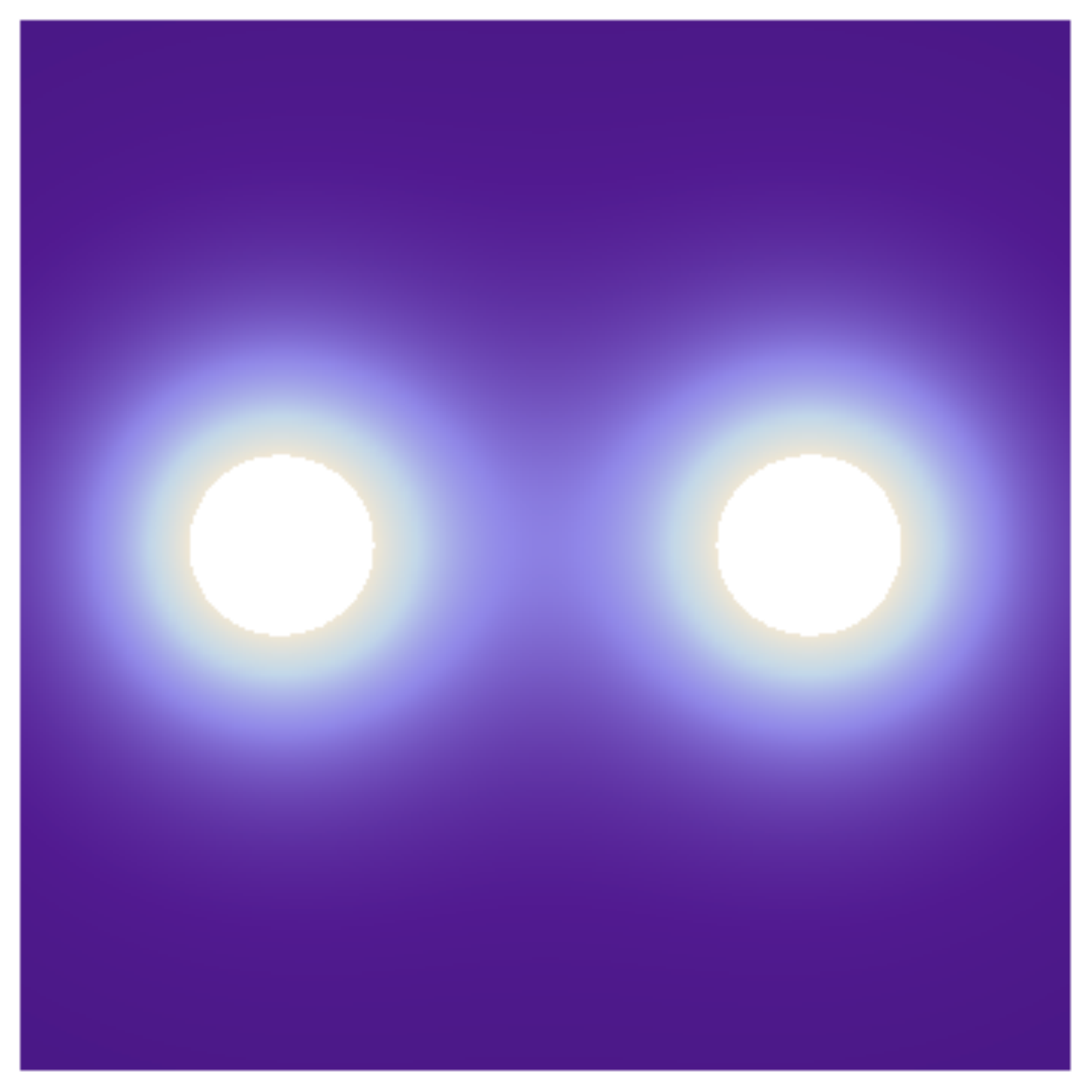}
\label{THGA2}}
\subfigure[\ $x_0=\lambda/(4\pi)$]{%
\includegraphics[width=0.3\linewidth]{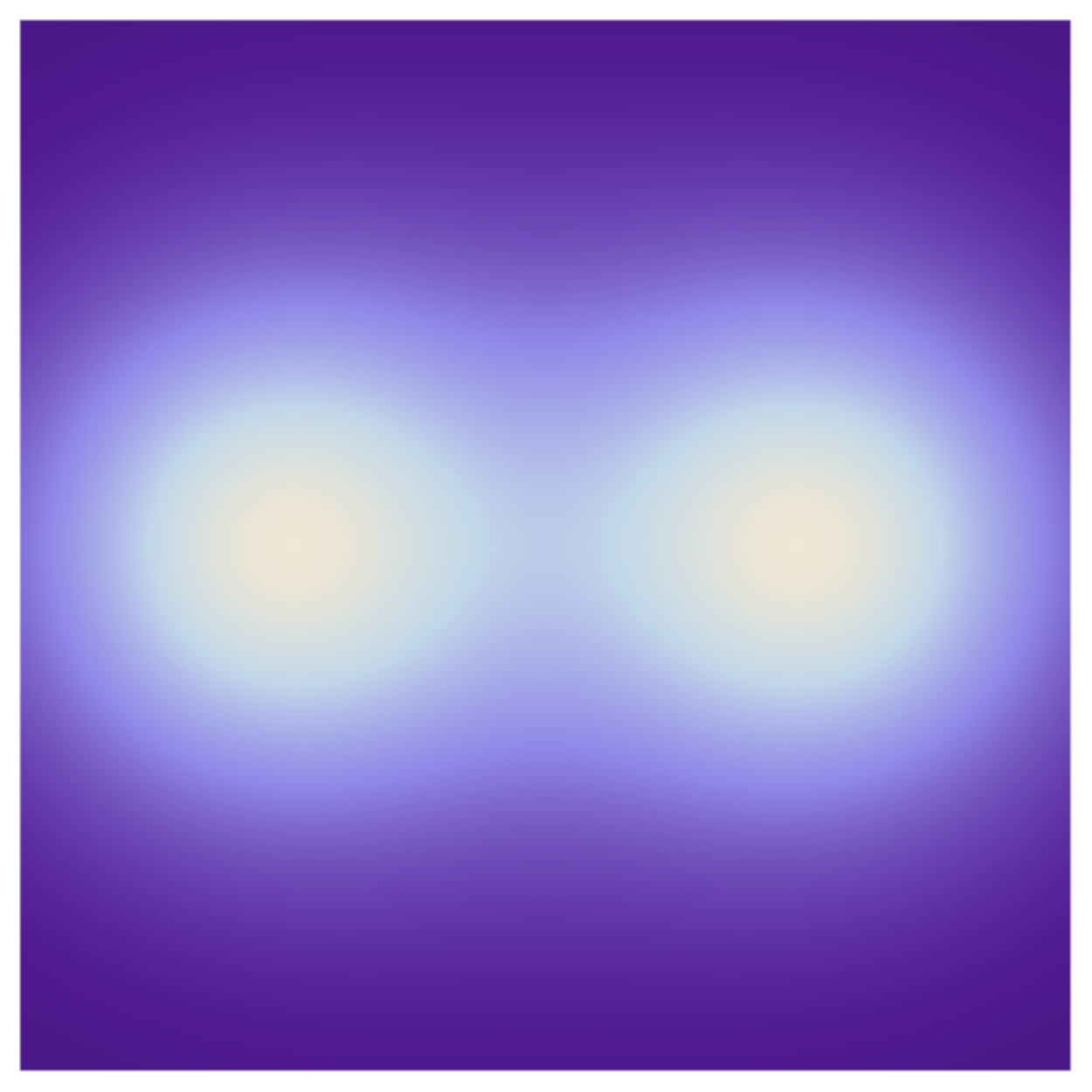}
\label{THGA2.5}}
\subfigure[\ $x_0=3\lambda/(10\pi)$]{%
\includegraphics[width=0.3\linewidth]{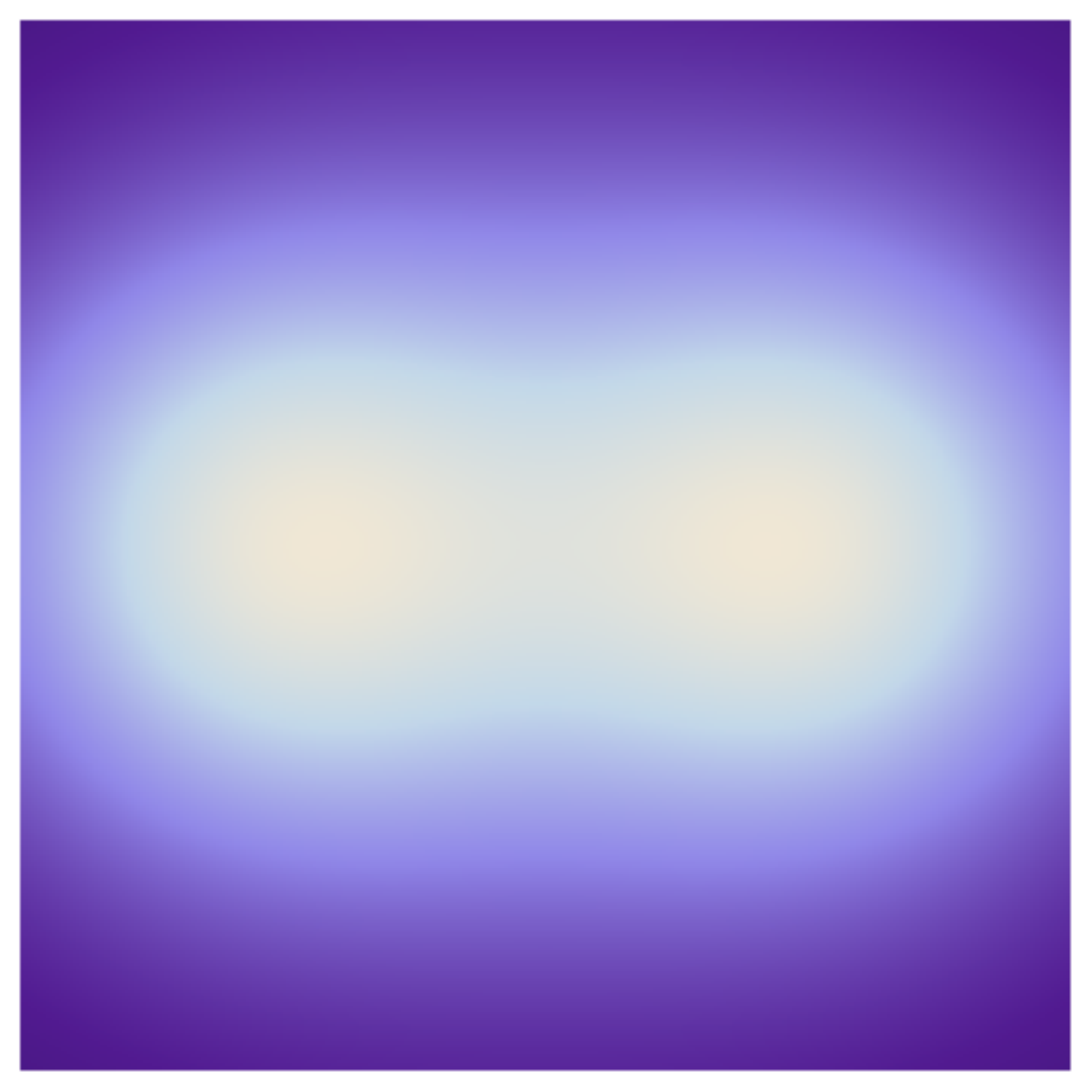}
\label{THGA3}}
\caption{(Color online) Apertureless NSOM images of the far-field intensity at frequency $3\Omega$ in THG.
The susceptibility of the tip is $\eta^{(3)}_{1111}E_0/\hat{\eta}^{(1)}=0.2$. 
Images are shown in the planes $x=x_0$ as indicated.
The field of view of each image is
$3\lambda/(5\pi) \times 3\lambda/(5\pi)$.}
\label{THGA}
\end{figure}

\begin{figure}
\includegraphics[width=0.8\linewidth]{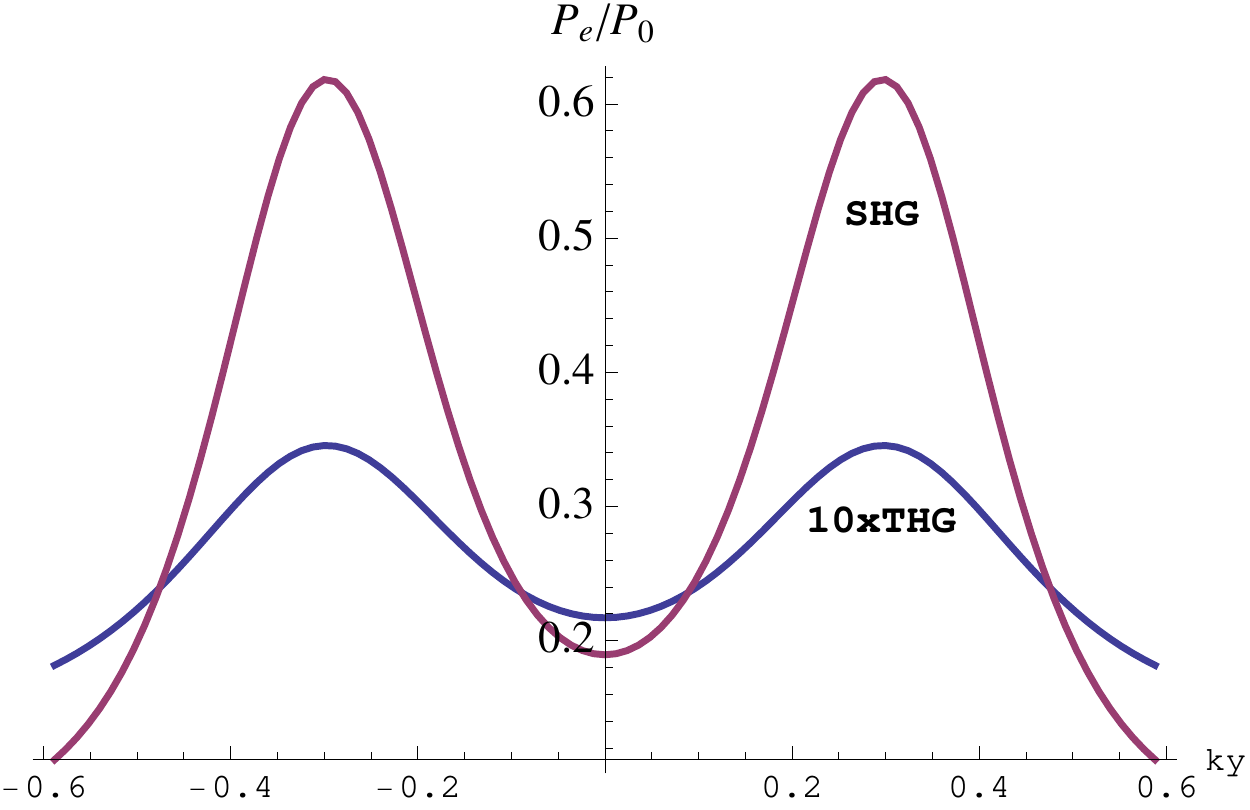}
\caption{(Color online) Extinguished power $P_e/P_0$ along the line defined by $x=\lambda/(5\pi)$ and $z=0$, which
corresponds to the closest scanning plane. Graphs are shown for SHG and THG.
{Here $P_0=a^2cE_0^2$.}}
\label{PeComp}
\end{figure}

\begin{figure}
\includegraphics[width=0.8\linewidth]{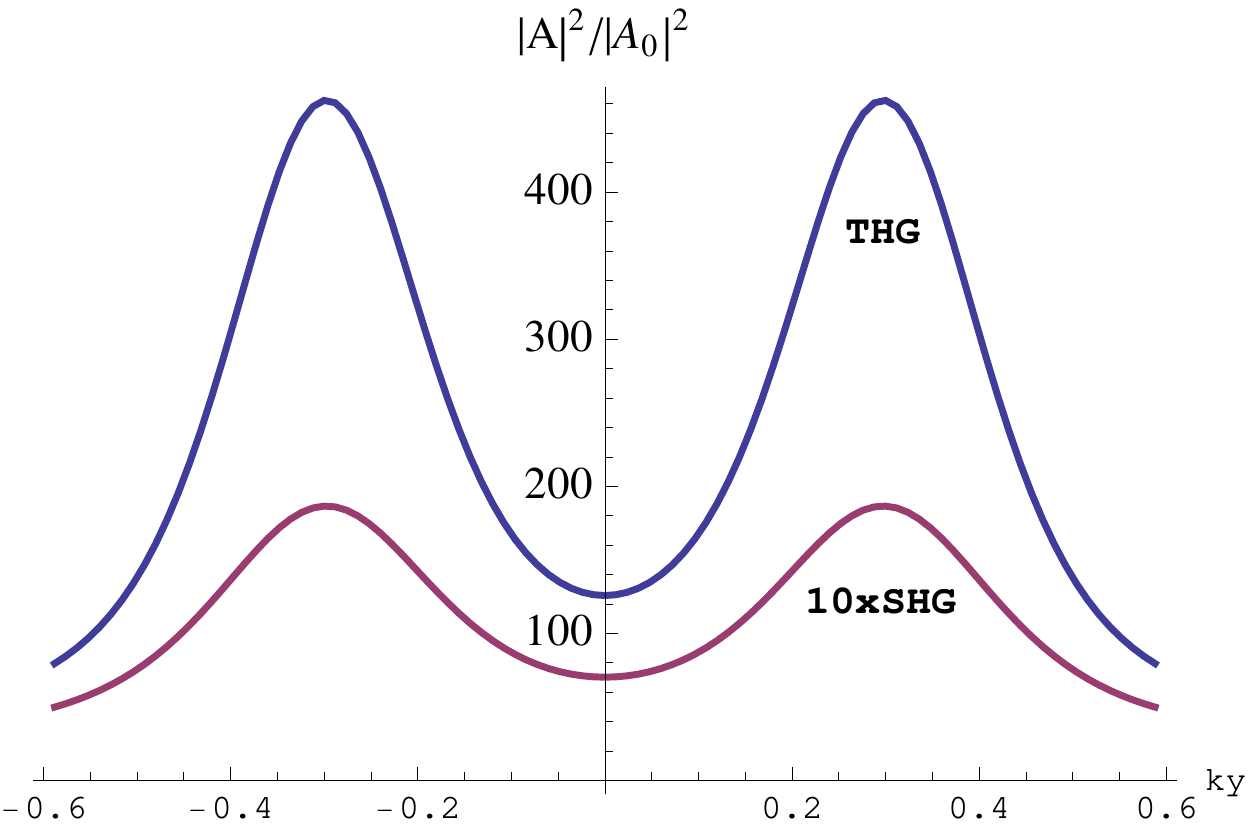}
\caption{(Color online) Far-field intensity along the line defined by $x=\lambda/(5\pi)$ and $z=0$, which
corresponds to the closest scanning plane. Graphs are shown for SHG and THG. 
Here $A_0=aE_0$.}
\label{AComp}
\end{figure}

\appendix
\section{Conservation of Energy}

Here we show that for nondispersive media,  the analog of the Manley-Rowe relations hold~\cite{Boyd}. That is, if the susceptibilities $\chi_{ij}^{(1)}$, $\chi_{ijk}^{(2)}$ and $\chi_{ijkl}^{(3)}$ are purely real, then $\nabla \cdot \bar{\textbf{S}} =0$. We treat the cases of quadratic and cubic nonlinearity separately. Note that for an incident field
consisting of a sum of a finite number of frequencies, the integral in (\ref{Poynting t}) becomes a sum.

\subsection{Quadratic Nonlinearity}
 
We begin by inserting the quadratic polarization (\ref{quadratic polarization}) into the statement of energy conservation (\ref{div current}). We then have
\begin{eqnarray}
\nabla\cdot\textbf{S}(\textbf{r},\omega)&=&\frac{\omega}{8\pi^3}\text{Im}\left(E^*_i(\textbf{r},\omega)(\chi^{(1)}_{ij}(\textbf{r};\omega)E_j(\omega)\right)\notag\\
&+&\frac{\omega}{8\pi^3}\text{Im}\left(\sum_{\omega_1+\omega_2=\omega}\chi^{(2)}_{ijk}(\textbf{r};\omega_1,\omega_2)E^*_i(\textbf{r},\omega)E_j(\textbf{r},\omega_1)E_k(\textbf{r},\omega_2)\right)\ .
\end{eqnarray}
Making use of (\ref{Poynting t}), the time-averaged divergence of energy current is 
\begin{eqnarray}
\label{energy flux}
\nabla\cdot\bar{\textbf{S}}(\textbf{r})&=&\sum_{\omega}\frac{\omega}{8\pi^3}\text{Im}(\chi^{(1)}_{ij}(\textbf{r};\omega)E_i^*(\textbf{r},\omega)E_j(\textbf{r},\omega))\notag\\
&&+\sum_{\omega_1+\omega_2=\omega}\frac{\omega}{8\pi^3}\text{Im}(\chi^{(2)}_{ijk}(\textbf{r};\omega_1,\omega_2)E^*_i(\textbf{r},\omega)E_j(\textbf{r},\omega_1)E_k(\textbf{r},\omega_2))\ .
\end{eqnarray}

The first sum is zero since $\chi^{(1)}_{ij}(\textbf{r};\omega)E_i^*(\textbf{r},\omega)E_j(\textbf{r},\omega)$ is real for each $\omega$, due to the permutation symmetry of $\chi^{(1)}_{ij}$.
The second sum vanishes as a consequence of the overall permutation symmetry of $\chi^{(2)}_{ijk}$
and the constraint $\omega_1+\omega_2=\omega$.

\subsection{Cubic Nonlinearity}

For cubic nonlinearity we have 
\begin{eqnarray}
\label{cubic energy flux}
\nabla\cdot\bar{\textbf{S}}(\textbf{r})&&=\sum_{\omega}\frac{\omega}{8\pi^3}\text{Im}(\chi^{(1)}_{ij}(\textbf{r};\omega)\textbf{E}_i^*(\textbf{r},\omega)\textbf{E}_j(\textbf{r},\omega))\notag\\
&&+\sum_{\omega_1+\omega_2+\omega_3=\omega}\frac{\omega}{8\pi^3}\text{Im}(\chi^{(3)}_{ijkl}(\textbf{r};\omega_1,\omega_2,\omega_3)\textbf{E}^*_i(\textbf{r},\omega)\textbf{E}_j(\textbf{r},\omega_1)\textbf{E}_k(\textbf{r},\omega_2)\textbf{E}_l(\textbf{r},\omega_3))
\end{eqnarray}

The first sum is zero by the same argument as above. The second sum can be shown to be zero using the overall permutation symmetry of $\chi^{(3)}_{ijkl}$ and the constraint $\omega_1+\omega_2+\omega_3=\omega$. 

\section{Calculation of local fields}
Here we calculate the local fields of small scatterers for both second harmonic generation and the Kerr effect.
\subsection{Second harmonic generation}
\label{local_fields_quadratic}
{
We now calculate the local fields $E_i(\textbf{0},\Omega)$ and $E_i(\textbf{0},2\Omega)$. To proceed, we set $\textbf r=0$ in (\ref{SHG int1}) and (\ref{SHG int2}) and thus obtain
\begin{eqnarray}
\label{local 1}
E_i(\textbf{0},\Omega)&=&E_{inc,i}(\textbf{0},\Omega)+k^2(\Omega)\eta_{jk}^{(1)}\int_{|\textbf{r}'|\le a} d^3r'G_{ij}(\textbf{0},\textbf{r}';\Omega)E_k(\textbf{r}',\Omega)\notag\\
&+&2k^2(\Omega)\eta_{jkl}^{(2)}\int_{|\textbf{r}'|\le a} d^3r'G_{ij}(\textbf{0},\textbf{r}';\Omega)E_k(\textbf{r}',\Omega)E_l^*(\textbf{r}',2\Omega) \ , \\
E_i(\textbf{0},2\Omega)&=&k^2(2\Omega)\eta_{jk}^{(1)}\int_{|\textbf{r}'|\le a} d^3r'G_{ij}(\textbf{0},\textbf{r}';2\Omega)E_k(\textbf{r}',2\Omega)\notag\\
&+&k^2(\Omega)\eta_{jkl}^{(2)}\int_{|\textbf{r}'|\le a} d^3r'G_{ij}(\textbf{0},\textbf{r}';2\Omega)E_k(\textbf{r}',\Omega)E_l(\textbf{r}',\Omega))\ .
\label{local 2}
\end{eqnarray}
Next, we use the fact that for a function $g_j$
\begin{eqnarray}
\label{Int}
\nonumber
\int_{|\textbf{r}'|\le a} d^3r'G_{ij}(\textbf{0},\textbf{r}';\omega)g_j(\textbf{r}',\omega)=\frac{4\pi a^3}{3}\left(\frac{1}{a}+i\frac{2}{3}k(\omega)-\frac{1}{a^3k^2(\omega)}+O(k^2(\omega)a)\right)g_i(\textbf{0},\omega)\left(1+O(k(\omega)a) \right) \ , \\
\end{eqnarray}
which is derived in Appendix~\ref{integral}. 
We then find that (\ref{local 1}) and (\ref{local 2}) lead to a system of equations for the local fields which are of the form
\begin{eqnarray}
\label{nonlinear_alg_1}
E_i(\textbf{0},\Omega)&=&E_{inc,i}(\textbf{0},\Omega)+\frac{4\pi}{3}k^2(\Omega)a^3G_R(\Omega)(\eta_{ij}^{(1)}E_j(\textbf{0},\Omega)+2\eta_{ijk}^{(2)}E_j(\textbf{0},2\Omega)E_k^*(\textbf{0},\Omega))\ , \\
E_i(\textbf{0},2\Omega)&=&\frac{4\pi}{3}k^2(2\Omega)a^3G_R(2\Omega)(\eta_{ij}^{(1)}E_j(\textbf{0},2\Omega)+\eta_{ijk}^{(2)}E_j(\textbf{0},\Omega)E_k(\textbf{0},\Omega))\ ,
\label{nonlinear_alg_2}
\end{eqnarray}
where
\begin{eqnarray}
\label{ren G}
G_R(\omega)=\frac{1}{a}+i\frac{2}{3}k(\omega)-\frac{1}{a^3k^2(\omega)} \ .
\end{eqnarray}}

{
The above is a set of nonlinear algebraic equations which we solve perturbatively. To proceed, we introduce a parameter $\epsilon$ to scale the nonlinear terms in (\ref{nonlinear_alg_1}) and (\ref{nonlinear_alg_2}):
\begin{eqnarray}
\label{nonlinear_alg_1_scaled}
E_i(\textbf{0},\Omega)&=&E_{inc,i}(\textbf{0},\Omega)+\frac{4\pi}{3}k^2(\Omega)a^3G_R(\Omega)(\eta_{ij}^{(1)}E_j(\textbf{0},\Omega)+2\epsilon\eta_{ijk}^{(2)}E_j(\textbf{0},2\Omega)E_k^*(\textbf{0},\Omega))\ , \\
E_i(\textbf{0},2\Omega)&=&\frac{4\pi}{3}k^2(2\Omega)a^3G_R(2\Omega)(\eta_{ij}^{(1)}E_j(\textbf{0},2\Omega)+\epsilon\eta_{ijk}^{(2)}E_j(\textbf{0},\Omega)E_k(\textbf{0},\Omega))\ .
\label{nonlinear_alg_2_scaled}
\end{eqnarray}
We then introduce asymptotic expansions for the fields of the form
\begin{eqnarray}
\label{asymptotic_1}
E_i(0,\Omega) &=& E_i^{(0)}(0,\Omega) + \epsilon E_i^{(1)}(0,\Omega) + \epsilon^2E_i^{(2)}(0,\Omega)
+ \cdots \ , \\
E_i(0,2\Omega) &= & E_i^{(0)}(0,2\Omega) + \epsilon E_i^{(1)}(0,2\Omega) + \epsilon^2E_i^{(2)}(0,2\Omega)
+ \cdots \ .
\label{asymptotic_2}
\end{eqnarray}}

{
For simplicity, we consider
the case of isotropic $\eta^{(1)}$ and $\eta^{(2)}$ obeying permutation symmetry.
That is, $\eta_{ij}^{(1)}=\eta^{(1)}\delta_{ij}$ and $\eta^{(2)}_{111}=\eta^{(2)}$, 
with the other $\eta^{(2)}_{ijk}$ vanishing.
We also assume that the incident field $\textbf{E}_{inc}$ points in the $x$-direction and the direction of observation $\hat{\textbf{s}}$ is taken to be in the $z$-direction. To simplify the notation,  we set $\Omega_1=\Omega$, $\Omega_2=2\Omega$, and write $(E_i)_j=E_j(\textbf{0},\Omega_i)$, $k_i=\Omega_i/c$, $(E_{inc})_i=E_{inc,i}(\textbf{0},\Omega)$ and $G_{Ri}=G_R(\Omega_i)$. Then (\ref{nonlinear_alg_1_scaled}) and (\ref{nonlinear_alg_2_scaled})
become
\begin{eqnarray}
(E_1)_i&=&(E_{inc})_i+\frac{4\pi}{3}k_1^2a^3G_{R1}(\eta^{(1)}(E_1)_i+2\epsilon\eta^{(2)}_{ijk}(E_1^*)_j(E_2)_k)\notag\\
(E_2)_i&=&\frac{4\pi}{3}k_2^2a^3G_{R2}(\eta^{(1)}(E_2)_i+\epsilon\eta^{(2)}_{ijk}(E_1)_j(E_1)_k)\ .
\end{eqnarray}
Next, we expand the fields $(E_i)_j$ according to (\ref{nonlinear_alg_1}) and (\ref{nonlinear_alg_2}) and collect like powers of $\epsilon$.
At  $O(1)$ we have that
\begin{eqnarray}
(E_1)_i^{(0)}&=&(E_{inc})_i+\frac{4\pi}{3}k_1^2a^3G_{R1}\eta^{(1)}(E_1)_i^{(0)}\notag\\
(E_2)_i^{(0)}&=&0\ .
\end{eqnarray}
Thus
\begin{eqnarray}
(E_1)_1^{(0)}&=&\frac{(E_{inc})_i}{1-\frac{4\pi}{3}k_1^2a^3G_{R1}\eta^{(1)}}\ .
\end{eqnarray}
At $O(\epsilon)$ we have
\begin{eqnarray}
(E_1)_i^{(1)}&=&\frac{4\pi}{3}k_1^2a^3G_{R1}\eta^{(1)}(E_1)_i^{(1)}\notag\\
(E_2)_i^{(1)}&=&\frac{4\pi}{3}k_2^2a^3G_{R2}(\eta^{(1)}(E_2)_i^{(1)}+\eta^{(2)}_{ijk}(E_1)_j^{(0)}(E_1)_k^{(0)})\ ,
\end{eqnarray}
which gives
\begin{eqnarray}
(E_2)_1^{(1)}&=&\frac{4\pi}{3}k_2^2a^3G_{R2}(\eta^{(1)}(E_2)_3^{(1)}+\eta^{(2)}_{111}(E_1)_1^{(0)}(E_1)_1^{(0)})\ .
\end{eqnarray}
Thus 
\begin{eqnarray}
(E_2)_1^{(1)}&=&\frac{\frac{4\pi}{3}k_2^2a^3G_{R2}\eta^{(2)}_{111}(E_1)_1^{(0)}(E_1)_1^{(0)}}{1-\frac{4\pi}{3}k_2^2a^3G_{R2}\eta^{(1)}}\ .
\end{eqnarray}
At $O(\epsilon^2)$ we obtain
\begin{eqnarray}
(E_1)_i^{(2)}&=&\frac{4\pi}{3}k_1^2a^3G_{R1}(\eta^{(1)}(E_1)_i^{(2)}+2\eta^{(2)}_{ijk}(E_1^*)_j^{(0)}(E_2)_k^{(1)})\ ,
\end{eqnarray}
which gives
\begin{eqnarray}
(E_1)_1^{(2)}&=&\frac{4\pi}{3}k_1^2a^3G_{R1}(\eta^{(1)}(E_1)_1^{(2)}+2\eta^{(2)}_{111}(E_2)_1^{(1)}(E_1^*)_1^{(0)}) \ .
\end{eqnarray}}
{
Thus
\begin{eqnarray}
(E_1)_1^{(2)}&=&\frac{\frac{4\pi}{3}k_1^2a^3G_{R1}2\eta^{(2)}_{111}(E_1^*)_1^{(0)}(E_2)_1^{(1)}}{1-\frac{4\pi}{3}k_1^2a^3G_{R1}\eta^{(1)}} \ .
\end{eqnarray}
We can now calculate the extinguished power $P_e$ from (\ref{OPT f}). We find that
up to the order $O(\epsilon^2)$
\begin{eqnarray}
\label{Pe_SHG}
P_e= \frac{8\Omega_1}{3} a^3 \text{Im}\left((\eta^{(1)}(E_1)_1^{(0)}+\eta^{(1)}(E_1)_1^{(2)}+2\eta^{(2)}_{111}(E_1^*)_1^{(0)}(E_2)_1^{(1)})(E_{inc})_1^*\right)\ .
\end{eqnarray}}

\subsection{Kerr effect}
\label{local_fields_cubic}
{
Here we calculate the local field $E_i(\textbf{0},\Omega)$. To proceed, we set $\textbf r=0$ in (\ref{Kerr int}) and thus obtain
\begin{eqnarray}
\label{local 3}
E_i(\textbf{0},\Omega)&=&E_{inc,i}(\textbf{0},\Omega)+k^2(\Omega)\eta_{jk}^{(1)}\int_{|\textbf{r}'|\le a} d^3r'G_{ij}(\textbf{0},\textbf{r}';\Omega)E_k(\textbf{r}',\Omega)\notag\\
&+&3k^2(\Omega) \eta^{(3)}_{jklm}\int_{|\textbf{r}'|\le a}  d^3 r' G_{ij}(\textbf{0},\textbf{r}';\Omega)E_k(\textbf{r}',\Omega)E_l(\textbf{r}',\Omega)E_m^*(\textbf{r}',\Omega)) \ .
\end{eqnarray}
We then find that (\ref{local 3}) leads to an equation for the local field which is of the form
\begin{eqnarray}
\label{nonlinear_alg_3}
E_i(\textbf{0},\Omega)=&&E_{inc,i}(\textbf{0},\Omega)+
\frac{4\pi}{3}k^2a^3(\Omega)G_{R}(\Omega)(\eta^{(1)}E_i(\textbf{0},\Omega)\notag\\
&&+3\eta^{(3)}_{ijkl}E^*_j(\textbf{0},\Omega)E_k(\textbf{0},\Omega)E_l(\textbf{0},\Omega))\ .
\end{eqnarray}}
{
The above is a nonlinear algebraic equation which we solve perturbatively. To proceed, we introduce a parameter $\epsilon$ to scale the nonlinear terms in (\ref{nonlinear_alg_3}):
\begin{eqnarray}
E_i(\textbf{0},\Omega)=&&E_{inc,i}(\textbf{0},\Omega)+
\frac{4\pi}{3}k^2a^3(\Omega)G_{R}(\Omega)(\eta^{(1)}E_i(\textbf{0},\Omega)\notag\\
&&+3\epsilon\eta^{(3)}_{ijkl}E^*_j(\textbf{0},\Omega)E_k(\textbf{0},\Omega)E_l(\textbf{0},\Omega))\ .
\label{nonlinear_alg_3_scaled}
\end{eqnarray}
We then introduce asymptotic expansions for the field of the form
\begin{eqnarray}
\label{asymptotic_3}
E_i(0,\Omega) &=& E_i^{(0)}(0,\Omega) + \epsilon E_i^{(1)}(0,\Omega) + \epsilon^2E_i^{(2)}(0,\Omega)
+ \cdots \ .
\end{eqnarray}}

{For simplicity, we consider
the case of isotropic $\eta^{(1)}$ and $\eta^{(3)}$ obeying the permutation symmetry. That is, $\eta_{ij}^{(1)}=\eta^{(1)}\delta_{ij}$ and $\eta_{1111}^{(3)}=\eta^{(3)}$, with all other $\eta_{ijkl}^{(3)}$ vanishing.
We also assume that the incident field $\textbf{E}_{inc}$ points in the $x$-direction and the direction of observation $\hat{\textbf{s}}$ is taken to be in the $z$-direction. To simplify the notation,  we write $(E)_j=E_j(\textbf{0},\Omega_i)$, $k=\Omega/c$ and $G_R=G_R(\Omega)$. Then (\ref{nonlinear_alg_3_scaled}) becomes
\begin{eqnarray}
(E)_i=(E_{inc})_i+\frac{4\pi}{3}k^2a^3G_R(\eta^{(1)}(E)_i+3\epsilon \eta^{(3)}_{ijkl}(E^*)_j(E)_k(E)_l)\ .
\end{eqnarray}
Next, we expand the fields $(E)_i$ according to (\ref{asymptotic_3}) and collect like powers of $\epsilon$.
At  $O(1)$ we have that
\begin{eqnarray}
(E)_i^{(0)}=(E_{inc})_i+\frac{4\pi}{3}k^2a^3G_R\eta^{(1)}(E)_i^{(0)}\ .
\end{eqnarray}
Thus
\begin{eqnarray}
(E)_1^{(0)}=\frac{E_{inc,i}}{1-\frac{4\pi}{3}k^2a^3G_R\eta^{(1)}}\ .
\end{eqnarray}
At $O(\epsilon)$ we have
\begin{eqnarray}
(E)_i^{(1)}=\frac{4\pi}{3}k^2a^3G_R(\eta^{(1)}(E)_i^{(1)}+3\eta^{(3)}_{ijkl}(E^*)_j^{(0)}(E)_k^{(0)}(E)_l^{(0)})\ ,
\end{eqnarray}
which gives
\begin{eqnarray}
(E)_i^{(1)}=\frac{4\pi}{3}k^2a^3G_R(\eta^{(1)}(E)_i^{(1)}+3\eta^{(3)}_{i111}(E^*)_1^{(0)}(E)_1^{(0)}(E)_1^{(0)})\ .
\end{eqnarray}
Thus
\begin{eqnarray}
(E)_i^{(1)}(\textbf{0})=\frac{\frac{4\pi}{3}k^2a^3G_R3\eta^{(3)}_{i111}(E^*)_1^{(0)}(0)(E)_1^{(0)}(0)(E)_1^{(0)}(0))}{1-\frac{4\pi}{3}k^2a^3G_R\eta^{(1)}} 
\end{eqnarray}
and
\begin{eqnarray}
(E)_1^{(1)}(\textbf{0})=\frac{\frac{4\pi}{3}k^2a^3G_R3\eta^{(3)}_{1111}(E^*)_1^{(0)}(0)(E)_1^{(0)}(0)(E)_1^{(0)}(0))}{1-\frac{4\pi}{3}k^2a^3G_R\eta^{(1)}}\ .
\end{eqnarray}
We can now calculate the extinguished power $P_e$ from (\ref{OPT f}). We find that
up to the order $O(\epsilon)$
\begin{eqnarray}
\label{Pe_kerr}
P_e= \frac{8\Omega_1}{3} a^3 \text{Im}\left((\eta^{(1)}(E)_1^{(0)}+\eta^{(1)}(E)_1^{(1)}+3\eta^{(3)}_{1111}(E^*)_1^{(0)}(E)_1^{(0)}(E)_1^{(0)})(E_{inc})_1^*\right)\ .
\end{eqnarray}}

\subsection{ Evaluation of the Integral (\ref{Int})}
\label{integral}

Here we show
\begin{eqnarray}
\int_{|\textbf{r}'|\le a} d^3r'G_{ij}(\textbf{0},\textbf{r}';\omega)=
\frac{4\pi a^3}{3}\left(\frac{1}{a}+i\frac{2}{3}k(\omega)-\frac{1}{a^3k^2(\omega)}+O(k(\omega)a)\right)\ .
\end{eqnarray}
For notational convenience, we put $k=k(\Omega)$. Then
\begin{eqnarray}
\int_{|\textbf{r}'|\le a} d^3r'G_{ij}(\textbf{0},\textbf{r}';\omega)
=\int_{|\textbf{r}|\le a} d^3r \left(\delta_{ij}+\frac{1}{k^2}\partial_i\partial_j\right)\frac{e^{ikr}}{r}\notag\\
\end{eqnarray}
The second term vanishes when $i\neq j$. Using the fact that
\begin{eqnarray}
\left(\nabla^2+k^2\right)\frac{e^{ikr}}{r}=-4\pi\delta(\textbf{r})\ ,
\end{eqnarray}
we have
\begin{eqnarray}
\int_{|\textbf{r}|\le a} d^3 r \frac{1}{k^2}\partial_i\partial_j\frac{e^{ikr}}{r}=\int_{|\textbf{r}|\le a} d^3 r \delta_{ij}\frac{1}{k^2}\frac{1}{3}\left(-k^2\frac{e^{ikr}}{r}-4\pi\delta(\textbf{r})\right) \notag \\
=-\delta_{ij}\left(\int_{|\textbf{r}|\le a} d^3 r \frac{e^{ikr}}{3r}+\frac{4\pi}{3k^2}\right)\ .
\end{eqnarray}
So
\begin{eqnarray}
\int_{|\textbf{r}'|\le a} d^3r'G_{ij}(\textbf{0},\textbf{r}';\omega)
= \delta_{ij}\left(\int_{|\textbf{r}|\le a} d^3 r \frac{2}{3}\frac{e^{ikr}}{r}-\frac{4\pi}{3k^2}\right)\ .
\end{eqnarray}
Since
\begin{eqnarray}
\int_{|\textbf{r}|\le a} d^3 r \frac{e^{ikr}}{r}=\frac{4\pi a^3}{3}\left(\frac{1}{a}+i\frac{2}{3}k-\frac{1}{a^3k^2}+O(ka)\right)\ ,
\end{eqnarray}
we obtain the required result.

\section{NSOM}
Here we derive the basic equations governing the NSOM experiments described in Section~\ref{ch:Application}. 

\subsection{Second Harmonic Generation}
The sample and the tip are taken to be small balls of radius $a$ centered at $\textbf{r}_0$, $\textbf{r}_1$ and  $\textbf{r}_2$. The corresponding susceptibilities are $\chi^{(1)}_{ij}(\textbf{r};\omega)=\hat{\eta}_{ij}^{(1)}$ for $|\textbf{r}-\textbf{r}_0|\le a$, $\chi^{(1)}_{ij}(\textbf{r};\omega)=\eta_{ij}^{(1)}$ for $|\textbf{r}-\textbf{r}_1|\le a$ and $|\textbf{r}-\textbf{r}_2|\le a$, and $\chi^{(2)}_{ijk}(\textbf{r};\omega)=\eta^{(2)}_{ijk}$ for $|\textbf{r}|\le a$. In this setting, the solutions to the wave equations of SHG (\ref{E1}) and (\ref{E2}) are
\begin{eqnarray}
\label{SHGN1}
E_i(\textbf{r},\Omega)&=&E_{inc,i}(\textbf{r},\Omega)+k^2(\Omega)\hat{\eta}_{jk}^{(1)}\int_{|\textbf{r}'-\textbf{r}_0|\le a} d^3 r'G_{ij}(\textbf{r},\textbf{r}';\Omega)E_k(\textbf{r}',\Omega)\notag\\
&+&k^2(\Omega)\eta_{jk}^{(1)}\int_{|\textbf{r}'-\textbf{r}_1|\le a} d^3 r'G_{ij}(\textbf{r},\textbf{r}';\Omega)E_k(\textbf{r}',\Omega)+k^2(\Omega)\eta_{jk}^{(1)}\int_{|\textbf{r}'-\textbf{r}_2|\le a} d^3 r'G_{ij}(\textbf{r},\textbf{r}';\Omega)E_k(\textbf{r}',\Omega)\notag\\
&+&2k^2(\Omega)\eta_{jkl}^{(2)}\int_{|\textbf{r}'-\textbf{r}_0|\le a} d^3 r' G_{ij}(\textbf{r},\textbf{r}';\Omega)E_k(\textbf{r}',2\Omega)E_l^*(\textbf{r}',\Omega) \ , \\
E_i(\textbf{r},2\Omega)&=&k^2(2\Omega)\hat{\eta}_{jk}^{(1)}\int_{|\textbf{r}'-\textbf{r}_0|\le a} d^3r' G_{ij}(\textbf{r},\textbf{r}';\Omega)E_k(\textbf{r}',2\Omega)\notag\\
&+&k^2(2\Omega)\eta_{jk}^{(1)}\int_{|\textbf{r}'-\textbf{r}_1|\le a} d^3r' G_{ij}(\textbf{r},\textbf{r}';\Omega)E_k(\textbf{r}',2\Omega)+k^2(2\Omega)\eta_{jk}^{(1)}\int_{|\textbf{r}'-\textbf{r}_2|\le a} d^3r' G_{ij}(\textbf{r},\textbf{r}';\Omega)E_k(\textbf{r}',2\Omega)\notag\\\notag\\
&+&k^2(2\Omega)\eta_{jkl}^{(2)}\int_{|\textbf{r}'-\textbf{r}_0|\le a} d^3 r'G_{ij}(\textbf{r},\textbf{r}';\Omega)E_k(\textbf{r}',\Omega)E_l(\textbf{r}',\Omega) \ .
\label{SHGN2}
\end{eqnarray}
Using the asymptotic form of the Green's function given in (\ref{G asymp}), we find that the scattered fields are of the form
\begin{eqnarray}
E_i^s (\textbf{r},\Omega)&=& A_i(\textbf{r},\Omega) \frac{e^{ik(\Omega)r}}{r} \ , \\
E_i^s(\textbf{r},2\Omega)&=&A_i(\textbf{r},2\Omega)\frac{e^{ik(2\Omega)r}}{r} \ ,
\end{eqnarray}
where the scattering amplitudes are defined by
\begin{eqnarray}
A_i(\textbf{r},\Omega)&=&\frac{4\pi}{3}a^3(\delta_{ij}-\hat{r}_i\hat{r}_j)k^2(\Omega)(\hat{\eta}_{jk}^{(1)}E_k(\textbf{r}_0,\Omega)+2\eta_{jkl}^{(2)}E_k(\textbf{r}_0,2\Omega)E_l^*(\textbf{r}_0,\Omega)) e^{ik(\Omega)\hat{\textbf{r}}\cdot\textbf{r}_0}\notag\\
&+&\frac{4\pi}{3}a^3(\delta_{ij}-\hat{r}_i\hat{r}_j)k^2(\Omega)\eta_{jk}^{(1)}E_k(\textbf{r}_1,\Omega)e^{ik(\Omega)\hat{\textbf{r}}\cdot\textbf{r}_1}\notag\\
&+&\frac{4\pi}{3}a^3(\delta_{ij}-\hat{r}_i\hat{r}_j)k^2(\Omega)\eta_{jk}^{(1)}E_k(\textbf{r}_2,\Omega)e^{ik(\Omega)\hat{\textbf{r}}\cdot\textbf{r}_2}\ , \\
A_i(\textbf{r},2\Omega)&=&\frac{4\pi}{3}a^3(\delta_{ij}-\hat{r}_i\hat{r}_j)k^2(2\Omega)(\hat{\eta}_{jk}^{(1)}E_j(\textbf{r}_0,2\Omega)+\eta_{jkl}^{(2)}E_k(\textbf{r}_0,\Omega)E_l(\textbf{r}_0,\Omega))e^{ik(\Omega)\hat{\textbf{r}}\cdot\textbf{r}_0} \notag\\
&+&\frac{4\pi}{3}a^3(\delta_{ij}-\hat{r}_i\hat{r}_j)k^2(2\Omega)\eta_{jk}^{(1)}E_k(\textbf{r}_1,2\Omega)e^{ik(2\Omega)\hat{\textbf{r}}\cdot\textbf{r}_1}\notag\\
&+&\frac{4\pi}{3}a^3(\delta_{ij}-\hat{r}_i\hat{r}_j)k^2(2\Omega)\eta_{jk}^{(1)}E_k(\textbf{r}_2,2\Omega)e^{ik(2\Omega)\hat{\textbf{r}}\cdot\textbf{r}_2}\ .
\end{eqnarray}
Setting $\textbf{r}=\textbf{r}_0$, $\textbf{r}=\textbf{r}_1$ and $\textbf{r}=\textbf{r}_2$ in (\ref{SHGN1}) and (\ref{SHGN2}), and carrying out the indicated integrations we obtain
\begin{eqnarray}
E_i(\textbf{r}_0,\Omega)&=&E_{inc,i}(\textbf{r}_0,\Omega)+\frac{4\pi}{3}k^2(\Omega)a^3G_R(\Omega)(\hat{\eta}_{ij}^{(1)}E_j(\textbf{r}_0,\Omega)+2\eta_{ijk}^{(2)}E_j(\textbf{r}_0,2\Omega)E_k^*(\textbf{r}_0,\Omega))\ , \notag\\
&+&\frac{4\pi}{3}k^2(\Omega)a^3G_{ij}(\textbf{r}_1,\textbf{r}_0;\Omega)\eta_{jk}^{(1)}E_k(\textbf{r}_1,\Omega)+\frac{4\pi}{3}k^2(\Omega)a^3G_{ij}(\textbf{r}_2,\textbf{r}_0;\Omega)\eta_{jk}^{(1)}E_k(\textbf{r}_2,\Omega)\\
E_i(\textbf{r}_0,2\Omega)&=&\frac{4\pi}{3}k^2(2\Omega)a^3G_R(2\Omega)(\hat{\eta}_{ij}^{(1)}E_j(\textbf{r}_0,2\Omega)+\eta_{ijk}^{(2)}E_j(\textbf{r}_0,\Omega)E_k(\textbf{r}_0,\Omega))\notag\\
&+&\frac{4\pi}{3}k^2(2\Omega)a^3G_{ij}(\textbf{r}_1,\textbf{r}_0;2\Omega)\eta_{jk}^{(1)}E_k(\textbf{r}_1,2\Omega)+\frac{4\pi}{3}k^2(2\Omega)a^3G_{ij}(\textbf{r}_2,\textbf{r}_0;2\Omega)\eta_{jk}^{(1)}E_k(\textbf{r}_2,2\Omega)\ ,\notag\\
\end{eqnarray}
\begin{eqnarray}
E_i(\textbf{r}_1,\Omega)&=&E_{inc,i}(\textbf{r}_1,\Omega)+\frac{4\pi}{3}k^2(\Omega)a^3G_R(\Omega)\eta_{ij}^{(1)}E_j(\textbf{r}_1,\Omega)\ , \notag\\
&+&\frac{4\pi}{3}k^2(\Omega)a^3G_{ij}(\textbf{r}_1,\textbf{r}_0;\Omega)(\hat{\eta}_{jk}^{(1)}E_k(\textbf{r}_0,\Omega) +2\eta_{jkl}^{(2)}E_k(\textbf{r}_0,2\Omega)E_l^*(\textbf{r}_0,\Omega)))\notag\\
&+&\frac{4\pi}{3}k^2(\Omega)a^3G_{ij}(\textbf{r}_1,\textbf{r}_2;\Omega)\eta_{jk}^{(1)}E_k(\textbf{r}_2,\Omega)\\
E_i(\textbf{r}_1,2\Omega)&=&\frac{4\pi}{3}k^2(2\Omega)a^3G_R(2\Omega)\eta_{ij}^{(1)}E_j(\textbf{r}_1,2\Omega)\notag\\
&+&\frac{4\pi}{3}k^2(2\Omega)a^3G_{ij}(\textbf{r}_1,\textbf{r}_0;2\Omega)(\hat{\eta}_{jk}^{(1)}E_k(\textbf{r}_0,2\Omega)+\eta_{jkl}^{(2)}E_k(\textbf{r}_0,\Omega)E_l(\textbf{r}_0,\Omega))\notag\\
&+&\frac{4\pi}{3}k^2(2\Omega)a^3G_{ij}(\textbf{r}_1,\textbf{r}_2;2\Omega)\eta_{jk}^{(1)}E_k(\textbf{r}_2,2\Omega)\ ,
\end{eqnarray}

\begin{eqnarray}
E_i(\textbf{r}_2,\Omega)&=&E_{inc,i}(\textbf{r}_2,\Omega)+\frac{4\pi}{3}k^2(\Omega)a^3G_R(\Omega)\eta_{ij}^{(1)}E_j(\textbf{r}_2,\Omega)\ , \notag\\
&+&\frac{4\pi}{3}k^2(\Omega)a^3G_{ij}(\textbf{r}_2,\textbf{r}_0;\Omega)(\hat{\eta}_{jk}^{(1)}E_k(\textbf{r}_0,\Omega)+2\eta_{jkl}^{(2)}E_k(\textbf{r}_0,2\Omega)E_l^*(\textbf{r}_0,\Omega))\notag\\
&+&\frac{4\pi}{3}k^2(\Omega)a^3G_{ij}(\textbf{r}_2,\textbf{r}_1;\Omega)\eta_{jk}^{(1)}E_k(\textbf{r}_1,\Omega)\\
E_i(\textbf{r}_2,2\Omega)&=&\frac{4\pi}{3}k^2(2\Omega)a^3G_R(2\Omega)\eta_{ij}^{(1)}E_j(\textbf{r}_2,2\Omega)\notag\\
&+&\frac{4\pi}{3}k^2(2\Omega)a^3G_{ij}(\textbf{r}_2,\textbf{r}_0;2\Omega)(\hat{\eta}_{jk}^{(1)}E_k(\textbf{r}_0,2\Omega)+\eta_{jkl}^{(2)}E_k(\textbf{r}_0,\Omega)E_l(\textbf{r}_0,\Omega))\notag\\
&+&\frac{4\pi}{3}k^2(2\Omega)a^3G_{ij}(\textbf{r}_2,\textbf{r}_1;2\Omega)\eta_{jk}^{(1)}E_k(\textbf{r}_1,2\Omega)\ ,
\end{eqnarray}
Following the procedure indicated in Appendix B, the above equations can be solved perturbatively for the local fields.

\subsection{Third Harmonic Generation}
As above, the sample and the tip are small balls of radius $a$ centered at $\textbf{r}_0$, $\textbf{r}_1$ and  $\textbf{r}_2$. The susceptibilities are $\chi^{(1)}_{ij}(\textbf{r};\omega)=\hat{\eta}_{ij}^{(1)}$ for $|\textbf{r}-\textbf{r}_0|\le a$, $\chi^{(1)}_{ij}(\textbf{r};\omega)=\eta_{ij}^{(1)}$ for $|\textbf{r}-\textbf{r}_1|\le a$ and $|\textbf{r}-\textbf{r}_2|\le a$, and $\chi^{(3)}_{ijkl}(\textbf{r};\omega)=\eta^{(3)}_{ijkl}$ for $|\textbf{r}|\le a$.
We begin with the general cubic-nonlinear wave equations which are correct to order $\epsilon$:
\begin{eqnarray}
\label{N31}
\nabla\times\nabla\times\textbf{E}(\textbf{r},\Omega)-k^2(\Omega)\textbf{E}(\textbf{r},\Omega)=&&4\pi k^2(\Omega)( \chi_{ij}^{(1)}(\textbf{r},\Omega)E_j(\textbf{r},\Omega)\notag\\
&&+3\chi_{ijkl}^{(3)}(\textbf{r},\Omega,\Omega,-\Omega)E_j(\textbf{r},\Omega)E_k(\textbf{r},\Omega)E_l^*(\textbf{r},\Omega)) \ . 
\end{eqnarray}

\begin{eqnarray}
\label{N32}
\nabla\times\nabla\times\textbf{E}(\textbf{r},3\Omega)-k^2(3\Omega)\textbf{E}(\textbf{r},3\Omega)=&&4\pi k^2(3\Omega)( \chi_{ij}^{(1)}(\textbf{r},3\Omega)E_j(\textbf{r},3\Omega)\notag\\
&&+\chi_{ijkl}^{(3)}(\textbf{r},\Omega,\Omega,\Omega)E_j(\textbf{r},\Omega)E_k(\textbf{r},\Omega)E_l(\textbf{r},\Omega)) \ . 
\end{eqnarray}

It follows immediately from (\ref{soln}) that the solution to (\ref{N31}) and (\ref{N32}) is given by
\begin{eqnarray}
\label{solnN3}
E_i(\textbf{r},\Omega)&=&E_{inc,i}(\textbf{r},\Omega)+k^2(\Omega)\int d^3 r' \chi_{jk}^{(1)}(\textbf{r}',\Omega)G_{ij}(\textbf{r},\textbf{r}';\Omega)E_k(\textbf{r}',\Omega)\notag\\
&+& 3k^2(\Omega) \int d^3 r' \chi_{jklm}^{(3)}(\textbf{r}',\Omega,\Omega,-\Omega)G_{ij}(\textbf{r},\textbf{r}';\Omega)E_k(\textbf{r}',\Omega)E_l(\textbf{r}',\Omega)E_m^*(\textbf{r}',\Omega)) \ ,
\end{eqnarray}

\begin{eqnarray}
E_i(\textbf{r},3\Omega)&=&k^2(3\Omega)\int d^3 r' \chi_{jk}^{(1)}(\textbf{r}',3\Omega)G_{ij}(\textbf{r},\textbf{r}';3\Omega)E_k(\textbf{r}',3\Omega)\notag\\
&+& 3k^23(\Omega) \int d^3 r' \chi_{jklm}^{(3)}(\textbf{r}',\Omega,\Omega,\Omega)G_{ij}(\textbf{r},\textbf{r}';3\Omega)E_k(\textbf{r}',\Omega)E_l(\textbf{r}',\Omega)E_m(\textbf{r}',\Omega)) \ ,
\label{solnN4}
\end{eqnarray}

For the specific set up described at the beginning of this section, (\ref{solnN3}) and (\ref{solnN4}) become
\begin{eqnarray}
\label{N3 int1}
E_i(\textbf{r},\Omega)&=&E_{inc,i}(\textbf{r},\Omega)+k^2(\Omega)\hat{\eta}_{jk}^{(1)}\int_{|\textbf{r}'|\le a} d^3 r'G_{ij}(\textbf{r},\textbf{r}';\Omega)E_k(\textbf{r}',\Omega)\notag\\
&+&k^2(\Omega)\eta_{jk}^{(1)}\int_{|\textbf{r}'-\textbf{r}_1|\le a} d^3 r'G_{ij}(\textbf{r},\textbf{r}';\Omega)E_k(\textbf{r}',\Omega)\notag\\
&+&k^2(\Omega)\eta_{jk}^{(1)}\int_{|\textbf{r}'-\textbf{r}_2|\le a} d^3 r'G_{ij}(\textbf{r},\textbf{r}';\Omega)E_k(\textbf{r}',\Omega)\notag\\
&+&3k^2(\Omega) \eta^{(3)}_{jklm}\int_{|\textbf{r}'|\le a}  d^3 r' G_{ij}(\textbf{r},\textbf{r}';\Omega)E_k(\textbf{r}',\Omega)E_l(\textbf{r}',\Omega)E_m^*(\textbf{r}',\Omega))\ .
\end{eqnarray}

\begin{eqnarray}
\label{N3 int2 1}
E_i(\textbf{r},3\Omega)&=&k^2(3\Omega)\hat{\eta}_{jk}^{(1)}\int_{|\textbf{r}'|\le a} d^3 r'G_{ij}(\textbf{r},\textbf{r}';3\Omega)E_k(\textbf{r}',3\Omega)\notag\\
&+&k^2(3\Omega)\eta_{jk}^{(1)}\int_{|\textbf{r}'-\textbf{r}_1|\le a} d^3 r'G_{ij}(\textbf{r},\textbf{r}';3\Omega)E_k(\textbf{r}',3\Omega)\notag\\
&+&k^2(3\Omega)\eta_{jk}^{(1)}\int_{|\textbf{r}'-\textbf{r}_2|\le a} d^3 r'G_{ij}(\textbf{r},\textbf{r}';3\Omega)E_k(\textbf{r}',3\Omega)\notag\\
&+&k^2(3\Omega) \eta^{(3)}_{jklm}\int_{|\textbf{r}'|\le a}  d^3 r' G_{ij}(\textbf{r},\textbf{r}';3\Omega)E_k(\textbf{r}',\Omega)E_l(\textbf{r}',\Omega)E_m(\textbf{r}',\Omega))\ .
\label{N3 int2 2}
\end{eqnarray}
Using the asymptotic form of the Green's function given in (\ref{G asymp}), we find that the scattered field is of the form
\begin{eqnarray}
E_i^s (\textbf{r},\Omega)&=& A_i(\textbf{r},\Omega) \frac{e^{ik(\Omega)r}}{r} \\
E_i^s (\textbf{r},3\Omega)&=& A_i(\textbf{r},3\Omega) \frac{e^{ik(3\Omega)r}}{r} \ ,
\end{eqnarray}
where the scattering amplitude is defined by
\begin{eqnarray}
\label{N3 amp1}
A_i(\textbf{r},\Omega)&=&\frac{4\pi}{3}a^3(\delta_{ij}-\hat{r}_i\hat{r}_j)k^2(\Omega)(\hat{\eta}_{jk}^{(1)}E_k(\textbf{r}_0,\Omega)+3\eta_{jklm}^{(3)}E_k(\textbf{r}_0,\Omega)E_l(\textbf{r}_0,\Omega)E_m^*(\textbf{r}_0,\Omega))\notag\\
&+&\frac{4\pi}{3}a^3(\delta_{ij}-\hat{r}_i\hat{r}_j)k^2(\Omega)\eta_{jk}^{(1)}E_k(\textbf{r}_1,\Omega)e^{ik(\Omega)\hat{\textbf{r}}\cdot\textbf{r}_1}\notag\\
&+&\frac{4\pi}{3}a^3(\delta_{ij}-\hat{r}_i\hat{r}_j)k^2(\Omega)\eta_{jk}^{(1)}E_k(\textbf{r}_2,\Omega)e^{ik(\Omega)\hat{\textbf{r}}\cdot\textbf{r}_2}
\end{eqnarray}
\begin{eqnarray}
\label{N3 amp2}
A_i(\textbf{r},3\Omega)&=&\frac{4\pi}{3}a^3(\delta_{ij}-\hat{r}_i\hat{r}_j)k^2(3\Omega)(\hat{\eta}_{jk}^{(1)}E_k(\textbf{r}_0,3\Omega)+\eta_{jklm}^{(3)}E_k(\textbf{r}_0,\Omega)E_l(\textbf{r}_0,\Omega)E_m(\textbf{r}_0,\Omega))\notag\\
&+&\frac{4\pi}{3}a^3(\delta_{ij}-\hat{r}_i\hat{r}_j)k^2(3\Omega)\eta_{jk}^{(1)}E_k(\textbf{r}_1,3\Omega)e^{ik(3\Omega)\hat{\textbf{r}}\cdot\textbf{r}_1}\notag\\
&+&\frac{4\pi}{3}a^3(\delta_{ij}-\hat{r}_i\hat{r}_j)k^2(3\Omega)\eta_{jk}^{(1)}E_k(\textbf{r}_2,3\Omega)e^{ik(3\Omega)\hat{\textbf{r}}\cdot\textbf{r}_2}
\end{eqnarray}

Setting $\textbf{r}=\textbf{r}_0$, $\textbf{r}=\textbf{r}_1$ and $\textbf{r}=\textbf{r}_2$ in (\ref{N3 int2 1}) and (\ref{N3 int2 2}), and carrying out the indicated integrations we find that
\begin{eqnarray}
E_i(\textbf{r}_0,\Omega)&=&E_{inc,i}(\textbf{r}_0,\Omega)+\frac{4\pi}{3}a^3k^2(\Omega)\hat{\eta}^{(1)}_{ij}G_R(\Omega)E_j(\textbf{r}_0,\Omega)\notag\\
&+&\frac{4\pi}{3}a^3k^2(\Omega)\eta^{(1)}_{jk}G_{ij}(\textbf{r}_0,\textbf{r}_1;\Omega)E_k(\textbf{r}_1,\Omega)\notag\\
&+&\frac{4\pi}{3}a^3k^2(\Omega)\eta^{(1)}_{jk}G_{ij}(\textbf{r},\textbf{r}_2;\Omega)E_k(\textbf{r}_2,\Omega)\notag\\
&+&
\frac{4\pi}{3}a^3k^2(\Omega)3 \eta^{(3)}_{ijkl}G_R(\Omega)E_j(\textbf{r}_0,\Omega)E_k(\textbf{r}_0,\Omega)E_l^*(\textbf{r}_0,\Omega))
\end{eqnarray}

\begin{eqnarray}
E_i(\textbf{r}_1,\Omega)&=&E_{inc,i}(\textbf{r}_1,\Omega)+\frac{4\pi}{3}a^3k^2(\Omega)\hat{\eta}^{(1)}_{jk}G_{ij}(\textbf{r}_1,\textbf{r}_0;\Omega)E_k(\textbf{r}_0,\Omega)\notag\\
&+&\frac{4\pi}{3}a^3k^2(\Omega)\eta^{(1)}_{ij}G_R(\textbf{r},\textbf{r}_1;\Omega)E_j(\textbf{r}',\Omega)\notag\\
&+&\frac{4\pi}{3}a^3k^2(\Omega)\eta^{(1)}_{jk}G_{ij}(\textbf{r}_1,\textbf{r}_2;\Omega)E_k(\textbf{r}_2,\Omega)\notag\\
&+&\frac{4\pi}{3}a^3k^2(\Omega)3 \eta^{(3)}_{jklm}G_{ij}(\textbf{r}_1,\textbf{r}_0;\Omega)E_k(\textbf{r}_0,\Omega)E_l(\textbf{r}_0,\Omega)E_m^*(\textbf{r}_0,\Omega))\ .
\end{eqnarray}

\begin{eqnarray}
E_i(\textbf{r}_2,\Omega)&=&E_{inc,i}(\textbf{r}_2,\Omega)+\frac{4\pi}{3}a^3k^2(\Omega)\hat{\eta}^{(1)}_{jk}G_{ij}(\textbf{r}_2,\textbf{r}_0;\Omega)E_k(\textbf{r}_0,\Omega)\notag\\
&+&\frac{4\pi}{3}a^3k^2(\Omega)\eta^{(1)}_{jk}G_{ij}(\textbf{r}_2,\textbf{r}_1;\Omega)E_k(\textbf{r}_1,\Omega)\notag\\
&+&\frac{4\pi}{3}a^3k^2(\Omega)\eta^{(1)}_{ij}G_R(\Omega)E_j(\textbf{r}_2,\Omega)\notag\\
&+&\frac{4\pi}{3}a^3k^2(\Omega)3 \eta^{(3)}_{jklm}G_{ij}(\textbf{r}_2,\textbf{r}_0;\Omega)E_k(\textbf{r}_0,\Omega)E_l(\textbf{r}_0,\Omega)E_m^*(\textbf{r}_0,\Omega))\ .
\end{eqnarray}

\begin{eqnarray}
E_i(\textbf{r}_0,3\Omega)&=&\frac{4\pi}{3}a^3k^2(3\Omega)\hat{\eta}^{(1)}_{ij}G_R(3\Omega)E_j(\textbf{r}_0,3\Omega)\notag\\
&+&\frac{4\pi}{3}a^3k^2(3\Omega)\eta^{(1)}_{jk}G_{ij}(\textbf{r}_0,\textbf{r}_1;3\Omega)E_k(\textbf{r}_1,3\Omega)\notag\\
&+&\frac{4\pi}{3}a^3k^2(3\Omega)\eta^{(1)}_{jk}G_{ij}(\textbf{r}_0,\textbf{r}_2;3\Omega)E_k(\textbf{r}_2,3\Omega)\notag\\
&+&\frac{4\pi}{3}a^3k^2(3\Omega)\eta^{(3)}_{jklm}G_R(3\Omega)E_k(\textbf{r}_0,\Omega)E_l(\textbf{r}_0,\Omega)E_m(\textbf{r}_0,\Omega))\ .
\end{eqnarray}

\begin{eqnarray}
E_i(\textbf{r}_1,3\Omega)&=&\frac{4\pi}{3}a^3k^2(3\Omega)\hat{\eta}^{(1)}_{jk}G_{ij}(\textbf{r}_1,\textbf{r}_0;3\Omega)E_k(\textbf{r}_0,3\Omega)\notag\\
&+&\frac{4\pi}{3}a^3k^2(3\Omega)\eta^{(1)}_{ij}G_R(3\Omega)E_j(\textbf{r}_1,3\Omega)\notag\\
&+&\frac{4\pi}{3}a^3k^2(3\Omega)\eta^{(1)}_{jk}G_{ij}(\textbf{r}_1,\textbf{r}_2;3\Omega)E_k(\textbf{r}_2,3\Omega)\notag\\
&+&\frac{4\pi}{3}a^3k^2(3\Omega)\eta^{(3)}_{jklm}G_{ij}(\textbf{r}_1,\textbf{r}_0;3\Omega)E_k(\textbf{r}_0,\Omega)E_l(\textbf{r}_0,\Omega)E_m(\textbf{r}_0,\Omega))\ .
\end{eqnarray}

\begin{eqnarray}
E_i(\textbf{r}_2,3\Omega)&=&\frac{4\pi}{3}a^3k^2(3\Omega)\hat{\eta}^{(1)}_{jk}G_{ij}(\textbf{r}_2,\textbf{r}_0;3\Omega)E_k(\textbf{r}_0,3\Omega)\notag\\
&+&\frac{4\pi}{3}a^3k^2(3\Omega)\eta^{(1)}_{jk}G_{ij}(\textbf{r}_2,\textbf{r}_1;3\Omega)E_k(\textbf{r}_1,3\Omega)\notag\\
&+&\frac{4\pi}{3}a^3k^2(3\Omega)\eta^{(1)}_{ij}G_R(3\Omega)E_k(\textbf{r}_2,3\Omega)\notag\\
&+& \frac{4\pi}{3}a^3k^2(3\Omega)\eta^{(3)}_{jklm}G_{ij}(\textbf{r}_2,\textbf{r}_0;3\Omega)E_k(\textbf{r}_0,\Omega)E_l(\textbf{r}_0,\Omega)E_m(\textbf{r}_0,\Omega))\ .
\end{eqnarray}
Following the procedure indicated in Appendix B, the above equations can be solved perturbatively for the local fields.

\acknowledgments
We thank the referee for helpful suggestions. This work was supported in part by the NSF grants DMR--1120923, DMS--1115574 and DMS--1108969.

\end{document}